\documentclass[iop,revtex4]{emulateapj}
\usepackage{lscape}
\usepackage{graphicx}
\usepackage{color}
\shortauthors{Aberasturi et al.}

\begin{document}

%% LaTeX will automatically break titles if they run longer than
%% one line. However, you may use \\ to force a line break if
%% you desire.

\title{Constraints on the binary Properties of mid to late T dwarfs from Hubble Space Telescope WFC3 Observations}
\thanks{Based on observations made with the NASA/ESA Hubble Space 
Telescope, obtained at the Space Telescope Science Institute, which is 
operated by the Association of Universities for Research in Astronomy 
Inc., under NASA contract NAS 5-26555. These observations are 
associated with programs 11631 and 11666.}

%% Use \author, \affil, and the \and command to format
%% author and affiliation information.
%% Note that \email has replaced the old \authoremail command
%% from AASTeX v4.0. You can use \email to mark an email address
%% anywhere in the paper, not just in the front matter.
%% As in the title, use \\ to force line breaks.
 \author{M. Aberasturi\altaffilmark{1}, A.J. Burgasser\altaffilmark{2}, A. Mora\altaffilmark{3}, E. Solano\altaffilmark{1}, E.  L. Mart\'{\i}n\altaffilmark{4},  I. N. Reid\altaffilmark{5} and D. Looper\altaffilmark{6}}

 \affil{$^{1}$Centro de Astrobiolog\'{\i}a (INTA-CSIC), Departamento de Astrof\'{\i}sica, PO Box 78, 28691 Villanueva de la Ca\~nada, Madrid, Spain}
 
\affil{$^{2}$Center for Astrophysics and Space Science, University of California San Diego, La Jolla, CA, 92093, USA}  %         
       
\affil{$^{3}$ESA$-$ESAC, Gaia SOC. P. O. Box 78 E-28691 Villanueva de la Ca\~nada, Madrid, Spain}  %          

\affil{$^{4}$Centro de Astrobiolog\'{\i}a (INTA-CSIC), Departamento de Astrof'sica. Carretera de Ajalvir km 4, E-28550 Torrej—n de Ardoz, Madrid, Spain}

\affil{$^{5}$Space Telescope Science Institute, 3700 San Martin Drive, Baltimore, MD 21218, USA}

\affil{$^{6}$Institute for Astronomy, University of Hawaii, 2680 Woodlawn Drive,
Honolulu, HI 96822}
%% Notice that each of these authors has alternate affiliations, which
%% are identified by the \altaffilmark after each name.  Specify alternate
%% affiliation information with \altaffiltext, with one command per each
%% affiliation.

%\altaffiltext{1}{Visiting Astronomer, Cerro Tololo Inter-American Observatory.
%CTIO is operated by AURA, Inc.\ under contract to the National Science
%Foundation.}
%\altaffiltext{2}{Society of Fellows, Harvard University.}
%\altaffiltext{3}{present address: Center for Astrophysics,
   % 60 Garden Street, Cambridge, MA 02138}
%\altaffiltext{4}{Visiting Programmer, Space Telescope Science Institute}
%\altaffiltext{5}{Patron, Alonso's Bar and Grill}

%% Mark off your abstract in the ``abstract'' environment. In the manuscript
%% style, abstract will output a Received/Accepted line after the
%% title and affiliation information. No date will appear since the author
%% does not have this information. The dates will be filled in by the
%% editorial office after submission.

\begin{abstract}
We used  HST/WFC3 observations of a sample of 26 nearby ($\le$20 pc) mid to late T dwarfs to search for cooler companions and measure  the multiplicity statistics of brown dwarfs.
Tightly-separated companions were searched for using a double-PSF fitting algorithm. We also compared our detection limits based on simulations to other prior T5+ brown dwarf binary programs. 
 No new wide or tight companions were identified, which is consistent with the number of known T5+ binary systems and the resolution limits of WFC3.  We use our results to add new  constraints to the binary fraction of T$-$type brown dwarfs. Modeling selection effects and adopting previously derived separation and mass ratio distributions, we find an upper limit total binary fraction of $<$16$\%$ and $<$25$\%$ assuming power law and flat mass ratio distributions respectively, which are consistent with previous results.  We also characterize a handful of targets around the L/T transition.  
\end{abstract}

%% Keywords should appear after the \end{abstract} command. The uncommented
%% example has been keyed in ApJ style. See the instructions to authors
%% for the journal to which you are submitting your paper to determine
%% what keyword punctuation is appropriate.


\section{Introduction}

Since their first theoretical prediction (\citealt{1963ApJ...137.1121K}; \citealt{1963PThPh..30..460H}),  brown dwarfs (BDs), objects with insufficient  mass to sustain stable hydrogen fusion,  have bridged the gap in temperature and mass between cold, very low mass stars (VLM; $M_{\odot}$ $\geq$ 0.075$M_{\odot}$) and the hottest, most massive giant planets ($M_{\odot}$ $\leq$ 0.013 $M_{\odot}$; \citealt{2000ApJ...542L.119C}; \citealt{2003ApJ...596..587B}). Since the first discoveries of brown dwarfs as companions to low luminosity sources, GD165B \citep{1988Natur.336..656B} and Gl229B (\citealt{1995Natur.378..463N}; \citealt{1995ApJ...444L.101G}; \citealt{1995Sci...270.1478O}) and free-floating systems (\citealt{1995Natur.377..129R}; \citealt{1997ApJ...491L.107R}; \citealt{1997A&A...327L..29M}), three new spectral classes have been introduced  to characterize  these low mass objects:  L dwarfs ($T_{eff }$ $\sim$ 2500 K $-$ 1500K, \citealt{1999ApJ...519..802K}, \citealt{1999AJ....118.2466M}), T  dwarfs ($T_{eff }$ $\sim$1500 K $-$500K, \citealt{2006ApJ...637.1067B}) and Y dwarfs ($T_{eff}$  $\leq$ 500 K; \citealt{2011arXiv1108.4678C}). With photospheres dominated by condensate clouds in L dwarfs and molecular gas species  in T and Y dwarfs, BDs allow us  to study planetary-like atmospheres without having to eliminate the glare of a host star. Thanks to  wide-field infrared and optical imaging  surveys such as the Two Micron All Sky Survey (2MASS, \citealt{2006AJ....131.1163S}) and DEep Near Infrared Survey of the southern sky (DENIS, \citealt{1997Msngr..87...27E}) and 
spectroscopic surveys as Sloan Digital Sky Survey (SDSS, \citealt{2000AJ....120.1579Y}), we  now know of  $\sim$1000 L and T type BDs belonging to  the field and young stellar clusters\footnote{See http://dwarfarchives.org for an up-to-date list of L, T and Y dwarfs.}. More recently,  deeper near and mid$-$infrared surveys like the UKIRT Infrared Deep Sky Survey (UKIDSS, \citealt{2007MNRAS.379.1599L}), the Canada-France Brown Dwarf Survey (CFBDS, \citealt{2008A&A...484..469D}) and the Wide-field Infrared Survey Explorer (WISE; \citealt{2010AJ....140.1868W}),  have allowed us  to explore the regime of late T and Y dwarfs. Most Y dwarfs have been identified as isolated field objects in WISE (e.g., \citealt{2011arXiv1108.4677K}; \citealt{2012ApJ...759...60T}), but discoveries such as WD 0806-661B \citep{luhman11} and CFBDSIR J1458+1013B \citep{2011ApJ...740..108L}, demonstrate the continued utility of companion searches.\\

Beyond discovery, searches for companions allow us to measure the statistics of multiple systems, which are particularly useful for testing formation scenarios for VLM stars and BDs.
While the binary fraction (BF) of solar$-$type stellar systems is $\sim$  65$\%$ \citep{1991A&A...248..485D} and early$-$type M stars $\sim$30$\%$$-$40$\%$ (\citealt{1997AJ....114.1992R}; \citealt{2004ASPC..318..166D}), measurement of multiplicity statistics for BDs have inferred fractions of 15$\%$$-$30$\%$ (\citealt{2007ApJ...668..492A}; \citealt{2007ApJ...659..655B}), indicating a mass dependence either in multiple formation or in the subsequent evolution of multiple systems. \\

The majority of the $\sim$ 100 VLM binary systems now known were uncovered with high angular resolution Hubble Space Telescope ($HST$) and/or ground$-$based adaptive optics (AO) imaging programs (\citealt{2006ApJ...647.1393L}; \citealt{2007AJ....133.2320S}). Those studies find that BD systems peak in mass ratio at  $M_{2}$/$M_{1}$  $\approx$ 0.8  with separations typically closer than $a$$<$20 AU \citep{2007ApJ...668..492A}. The statistics of VLM binaries have motivated new theories of BD formation, via turbulent fragmentation \citep{2009MNRAS.392..590B} or gravitational instability in circumstellar disks \citep{2009MNRAS.392..413S}. Other techniques such as astrometry and analysis of microlensing events are reaching the sensitivity required to detect BD binaries with low mass ratios, and even giant planets around BDs (\citealt{2010ApJ...710.1142B}, \citealt{2011A&A...532A..31R}; \citealt{2013A&A...556A.133S}; \citealt{2013ApJ...768..129C}). Additionally, other techniques such as astrometry, high precision radial velocities, microlensing and blended-light spectroscopy are reaching the sensitivity required to detect BD binaries with low mass ratios, and even giant planets around BDs (\citealt{2010ApJ...710.1142B}, \citealt{2011A&A...532A..31R}; \citealt{2013A&A...556A.133S}; \citealt{2013ApJ...768..129C}).\\

These studies can be advanced by increasing the population of known brown dwarf binary systems.  To do this, we undertook two parallel programs using the Wide Field Camara 3 (WFC3) installed on  the Hubble Space Telescope. Observations, sample composition and data reduction are described in Section \ref{observations}. In Section \ref{photometry} we present the photometric results and define new color/spectral type relations. In Section \ref{psffitting} we describe the results of point$-$spread function (PSF) fits to our sample and results. We also quantify the WFC3 detection limits which prove to be the limitation in our study. In section \ref{analysis}, we infer a bias corrected BD binary fraction through simulation, and compare these results with previous  surveys. Finally, in Section \ref{conclusions} we summarize the main results of our project.

\section{Observations}
\label{observations}
\subsection{Sample}
Our original sample consists of  37 nearby sources  (d$\leq$ 30 pc) identified as L or T dwarfs based on prior searches of  2MASS, DENIS, SDSS or UKIDSS.  Measurements of the infrared photometry (2MASS and MKO systems), proper motions (PMs) and distances for the whole sample are listed in Table \ref{maintable}. We have used the \cite{2012arXiv1201.2465D} absolute magnitude$-$SpT relation to estimate the photometric distances for six objects without parallax measurement. The sources were observed as part of two HST(WFC3) programs  (11631 and 11666) with  slightly different goals:
\begin{itemize}

\item   Program 11631 targeted  11 L and early T dwarfs, including one known resolved binary 2MASS J1520$-$4422AB  \citep{2007ApJ...658..557B} and one previously unreported L3 source, DENIS J1013$-$7842 (Looper et al in prep). We present additional information of these sources in Appendix \ref{2MASSJ1520$-$4422AB} and \ref{DENISJ1013$-$7842}, respectively. The 11631 program aimed to explore multiplicity across the L/T transition and was used here to estimate magnitude-color and color-near infrared spectral type (NIR SpT) relations. The observations were obtained between January 2010 and June 2011.

\item Program 11666 targeted  26 mid and late-T dwarfs (from T5 to T8.5) to search Y dwarf companions, with measured or estimated distances  $\leq$ 20 pc not previously observed by HST or ground-based AO programs, with the exception of two sources with insufficient data (SDSS J1346$-$0031 and 2MASS J0727+1710). The observations were obtained between November 2009 and October 2010. The sample includes two of the coolest objects  known at that time: ULAS J0034-0052 \citep{2007MNRAS.381.1400W} and ULAS J1238+0953 \citep{2008MNRAS.391..320B}. We included one previously unreported T6 dwarf, 2MASS J2237+7228  (see more details in Appendix \ref{2MASSJ2237+7228}). In  Fig.\ref{histogram} we compare our sample against the number of known T5+ dwarfs within 20 parsecs; our sample includes $\sim$29$\%$ of such systems. We consider this sample statistically representative to estimate the binary fraction (BF) for the mid-late T dwarfs. 
\end{itemize}

The programs were originally planned for Near Infrared Camera and Multi-Object Spectrometer (NICMOS/NIC1) given that instrument's demonstrated ability to identify cold BD companions (\citealt{2006ApJS..166..585B}, \citealt{2011A&A...525A.123S}), but an instrument failure forced the change to WFC3.

\begin{deluxetable*}{lccccccccccc}
\tabletypesize{\scriptsize}
%\rotate
\tablecaption{L and T dwarfs sample \label{maintable}}
\tablewidth{0pt}
\tablehead{
\colhead{Name}         &
\colhead{NIR$-$SpT}&
\colhead{$\alpha$ $(J2000)$}&
\colhead{$\delta$ $(J2000)$}&
\colhead{$J$}&
\colhead{$H$}&
\colhead{$K_{s}$}&
\colhead{$\mu_{\alpha}$$cos_{\delta}$}&
\colhead{$\mu_{\delta}$}&
\colhead{Distance}&
\colhead{$\pi$}&
\colhead{References}\\
\colhead{} &
\colhead{Literature}&
\colhead{hh:mm:ss }&
\colhead{hh:mm:ss }&
\colhead{(mag)}&
\colhead{(mag)}&
\colhead{(mag)}&
\colhead{(mas $yr^{1}$)}&
\colhead{(mas $yr^{1}$)}&
\colhead{(pc)}&
\colhead{(mas)}&
\colhead{} \\
\colhead{(1)}&
\colhead{(2)}&
\colhead{(3)}&
\colhead{(4)}&
\colhead{(5)}&
\colhead{(6)}&
\colhead{(7)}&
\colhead{(8)}&
\colhead{(9)}&
\colhead{(10)}&
\colhead{(11)}&
\colhead{(12)}\\
}
\startdata
\multicolumn{12}{c}{Program 11631}\\                                                 
\tableline\tableline    
Name    &    NIR$-$SpT     &$\alpha$ $(J2000)$      & $\delta$ $(J2000)$      & $J$                          & $H$           & $K_{s}$      &   $\mu_{\alpha}$$cos_{\delta}$  & $\mu_{\delta}$   & Distance    &     $\pi$           &    References   \\
              &    Literature            & hh:mm:ss                       & hh:mm:ss                      & (mag)                      & (mag)        &   (mag)        &  (mas $yr^{1}$)                              &    (mas $yr^{1}$) &(pc)              &     (mas)         &         \\                                        
(1)          & (2)                 &  (3)                                  &  (4)                                 & (5)                             &     (6)            &   (7)           & (8)                                                    &  (9)                         & (10)                &   (11)           &  (12) \\
\tableline 
 2MASSI J0340-6724   &   L7::$^{a}$ & 03:40:09.42  & -67:24:05.1  & 14.74 $\pm$0.03 &  13.59$\pm$0.03 & 12.93$\pm$ 0.03  &  $-$318.0$\pm$7.0  & 508.0$\pm$18.0       & 11.0$\pm$3.0  &    \nodata    & (1);(15) \\
 SDSS J0739+6615 	    &   T1.5+/-1     & 07:39:22.26  & +66:15:03.9 & 16.82 $\pm$0.13 & 16.00$\pm$0.10  & 15.83$\pm$  0.18 &  180.0$\pm$10. 0     & $-$77.0$\pm$26.0   & 34.0$\pm$4.0  &   \nodata    & (2);(15) \\
 DENIS J1013-7842    &   L3$^{a}$   & 10:13:25.88  & -78:42:55.3  & 13.84$\pm$0.03  & 12.74$\pm$0.03  & 12.03$\pm$0.03   &  \nodata                                &         \nodata                          &14.2$\pm$1.3$^{c}$ &    \nodata   & (21) \\
 2MASS J1122-3512    &  T2                & 11:22:08.26  & -35:12:36.3  & 15.02$\pm$0.04  & 14.36$\pm$0.05  & 14.38$\pm$  0.07 &  $-$150.0$\pm$40.0 & $-$250.0$\pm$30.0 & 15.0$\pm$1.0  &   \nodata  & (3);(15)  \\
 SDSS J1439+3042 	    & T2.5              & 14:39:45.95  & +30:42:21.0 &  17.22$\pm$0.23 &  $>$16.28             &  $>$15.88               &        \nodata                            &               \nodata                   & 29.9$\pm$7.5$^{c}$  &  \nodata  & (2)\\
SDSS J1511+0607	    & T2                & 15:11:14.66  & +06:07:43.1  & 15.88$\pm$0.02 & 15.14$\pm$0.02  &  14.52$\pm$ 0.10  & $-$255.6$\pm$7.1    & $-$238.0$\pm$7. 0  &18.0$\pm$3.0  &  36.7$\pm$6.4 &(23);(16) \\
 2MASS J1520-4422A & L1.5             & 15:20:02.30  & -44:22:41.9   & 13.22$\pm$0.03  & 12.36$\pm$0.03  &  11.89$\pm$ 0.03  &$-$630.0$\pm$30.0  & $-$370.0$\pm$20.0 & 19.0$\pm$1.0  & \nodata    & (4);(15);(20) \\
 2MASS J1520-4422B & L4.5             & 15:20:02.30  & -44:22:41.9   & 14.70$\pm$0.07  & 13.70$\pm$0.05  &  13.70$\pm$ 0.05  &$-$630.0$\pm$30.0  &  $-$370.0$\pm$20.0 & 19.0$\pm$1.0 &  \nodata  & (4);(15);(20)  \\  
 %                                            &                   &                        &                         &                                   &                                 &                                  &                                      &                                       &                           &        &                                                 \\               
\multicolumn{12}{c}{Program 11666} \\ 
\tableline
\tableline 
%                                                &                &                         &                         &                                     &                             &                                  &                                        &                                       &                                &               &                                                  \\
ULAS J0034-0052$^{b}$    &  T8.5    & 00:34:02.76  & -00:52:08.0     &  18.15$\pm$0.08  &   18.49$\pm$0.04   &  18.48$\pm$0.05     &   \nodata                            &           \nodata                    &  12.6$\pm$0.6    & 79.60$\pm$3.80    &(5);(16) \\
HD3651B$^{b}$                   &   T7.5   & 00:39:18.61  & +21:15:12.7    & 16.16$\pm$0.03      &  16.68$\pm$0.04    &  16.87$\pm$0.        &  $-$461.1$\pm$0.7            &  $-$370.9$\pm$0.7       &  11.0$\pm$0.1     &  90.03$\pm$0.72  &(6);(17);(15) \\
2MASS J0050-3322            & T7        & 00:50:19.92  & -33:22:41.4    & 15.93$\pm$0.07       & 15.84$\pm$0.19     &  15.24 $\pm$0.19  & 1200.0$\pm$110.0  & 900.0$\pm$120.0  &  8.0$\pm$1.0        &   94.6$\pm$2.4       &(3);(15);(22) \\
SDSS J0325+0425           &  T5.5    & 03:25:53.11  & +04:25:40.0   & 16.25$\pm$0.14       &   $>$16.08                &  16.37$\pm$ 0.06  &  $-$163.7$\pm$5.8  &  $-$59.6$\pm$5.7  & 19.0$\pm$2.0      & 55.6$\pm$10.9     & (2);(16)  \\
2MASS J0407+1514  	  &  T5       & 04:07:08.94  & +15:14:55.4   & 16.06 $\pm$0.09      &  16.02$\pm$0.21    &  15.92$\pm$0.26    & 106.0$\pm$16.0       &$-$110.0$\pm$17.0&  17.0$\pm$2.0    &   \nodata                            & (7);(15) \\
2MASS J0510-4208  	  &  T5       & 05:10:35.32  & -42:08:08.2    & 16.22 $\pm$0.09      & 16.24$\pm$0.16     &  16.0$\pm$ 0.28      & 104.0$\pm$15.0      &580.0$\pm$21.0      &  18.0$\pm$2.0       &    \nodata                        &(8);(15)  \\
2MASSI J0727+1710 	  &  T7       & 07:27:19.07  & +17:09:52.2   & 15.60 $\pm$0.06     & 15.76$\pm$0.17     &  15.55$\pm$0.19    & 1046.0$\pm$4.0       &$-$767.0$\pm$3.0   & 9.1$\pm$0.2           & 110.14$\pm$2.34 & (9);(18) \\
2MASS J0729-3954  	  & T8pec  & 07:28:59.47  & -39:53:46.3     & 15.92 $\pm$0.08    &  15.98$\pm$0.18    &  $>$15.29                & $-$566.6$\pm$5.3   &1643.4$\pm$5.5        &   6.0$\pm$1.0         &126.3$\pm$8.3     &(8);(16) \\
2MASS J0741+2351  	  & T5         & 07:41:48.96  & +23:51:25.9    & 16.15$\pm$0.10     &   15.84$\pm$0.18   &  $>$15.85                & $-$243.0$\pm$13.0 & $-$143.0$\pm$14.0 &  18.0$\pm$2.0       &      \nodata                       &(10);(15) \\
2MASS J0939-2448            & T8         & 09:39:35.87  & -24:48:38.0     & 15.98 $\pm$0.11    &   15.80$\pm$0.15   &  $>$16.56                & 558.1$\pm$5.8          &$-$1030.5$\pm$5.6  & 10.0$\pm$2.0       & 196.0$\pm$10.4   &(3);(16)\\
2MASS J1007-4555  	  &  T5        & 10:07:32.99  & -45:55:13.3     & 15.65 $\pm$0.07    &  15.68$\pm$0.12    &  15.56 $\pm$0.23   & $-$723.5$\pm$3.4    &148.7$\pm$3.6          &  15.0$\pm$2.0      &71.0$\pm$5.2         & (8);(16) \\
2MASS J1114-2618            & T7.5      & 11:14:48.90  & -26:18:27.2     & 15.86$\pm$0.08     &   15.73$\pm$0.12  &  $>$16.11                 & $-$2927.2$\pm$7.0   & $-$374.2$\pm$7.2 & 10.0$\pm$2.0      &176.8$\pm$7.0        &(3);(16) \\
2MASS J1231+0847           & T5.5      & 12:31:46.74  & +08:47:22.3     & 15.57$\pm$0.07    & 15.31$\pm$0.11    &  15.22$\pm$0.19     & $-$1176.0$\pm$21.0 & $-$1043.0$\pm$21.0  &  12.0$\pm$1.0     &  \nodata                                  &(7);(15)\\
ULAS J1238+0953$^{b}$  & T8.5      & 12:38:28.57  & +09:53:51.3     & 18.95$\pm$0.02    & 19.20$\pm$0.02     &    \nodata                            &      \nodata                              &    \nodata                             &  18.5$\pm$4.3$^{c}$  & \nodata             &(11) \\
SDSS J1250+3925           	  & T4        & 12:50:11.67    & +39:25:47.9    & 16.54$\pm$0.11    & 16.18$\pm$0.18    & 15.05$\pm$  0.24   &   $-$15.0$\pm$80       & $-$828.0$\pm$11.0& 23.0$\pm$2.0    &     \nodata                                 & (2);(15) \\
SDSSP J13464-0031          & T6.5     & 13:46:46.04    & -00:31:51.3    & 16.00 $\pm$0.10   &15.46$\pm$0.12     &  15.77$\pm$  0.27  & $-$503.0$\pm$3.0      & $-$114.0$\pm$2.0   & 14.6$\pm$0.5    &    68.3$\pm$2.3       &(12);(19) \\
SDSS J1504+1027              & T7        & 15:04:11.74    & +10:27:16.8    &   17.03$\pm$0.23  & $>$16.90                 &   $>$17.02              & 373.8$\pm$ 7.9           &$-$322.5$\pm$7.7    & 15.9$\pm$2.5$^{c}$   & 52.5$\pm$7.1 & (2);(16)\\
SDSS J1628+2308  	            & T7       & 16:28:38.99   & +23:08:18.4     & 16.45$\pm$0.10     & 16.11$\pm$0.15    & 15.87$\pm$ 0.24   & 497.0$\pm$20.0         &$-$461.0$\pm$21.0  & 14.0$\pm$4.0    &      75.1$\pm$0.9      &(2);(15);(22) \\
2MASS J1754+1649           &  T5       & 17:54:54.56   & +16:49:18.1     &   15.81$\pm$0.07   & 15.65$\pm$0.13    &  15.55$\pm$0.16   & 113.5$\pm $9.1          & $-$141.4$\pm$9.2   & 14.3$\pm$1.3$^{c}$      &87.6$\pm$10.2  & (13) \\
SDSS J1758+4633              & T6.5    & 17:58:05.49   & +46:33:17.1     & 16.15 $\pm$0.08    &   16.25$\pm$0.21  & 15.46$\pm$  0.19   & 26.0$\pm$	15.0	        &594.0$\pm$16.0       & 12.0$\pm$2.0    &     71.0$\pm$1.9        &  (10);(15);(22) \\
2MASS J1828-4849            &  T5.5    &18:28:36.01    & -48:49:02.6      & 15.18 $\pm$0.06    & 14.91$\pm$0.07    &  15.18 $\pm$ 0.14  &231.4$\pm$10.5         &52.4$\pm$10.9          & 11.0$\pm$1.0    &83.7$\pm$7.7                & (7);(16)\\
2MASS J1901+4718           &  T5       & 19:01:05.89   & +47:18:09.9     &  15.86 $\pm$0.07   & 15.47$\pm$0.09   &  15.64 $\pm$ 0.29   & $-$110.0$\pm$20.0  & $-$360.0$\pm$20.0  &   15.0$\pm$2.0   &      \nodata                              &(7);(15)\\
SDSSJ 2124+0100              &  T5      & 21:24:14.02   & +01:00:02.7      & 16.03 $\pm$0.07   & 16.18$\pm$0.20   &  $>$16.14                 & 202.0$\pm$14.0          &   287.0$\pm$14.0      & 18.0$\pm$2.0   &      \nodata                               &(10);(15)\\
2MASS J2154+5942           & T5       & 21:54:32.98   & +59:42:14.4      & 15.66$\pm$0.07    & 15.76$\pm$0.17   &  $>$15.34                  & $-$182.0$\pm$9.0	& $-$445.0$\pm$17.0   &  10.0$\pm$1.0  &      \nodata                            & (8);(15)\\
2MASS J2237+7228           &  T6$^{a}$      & 22:37:20.47   & +72:28:35.3      &  15.76$\pm$0.07   & 15.94$\pm$0.21   & $>$15.99                   &  $-$73.0$\pm$2.0       &      $-$116.0$\pm$2.0  &   13.0$\pm$2.0 &      \nodata                          & (8)\\
2MASS J2331-4718            & T5       & 23:31:23.84  & -47:18:28.2        & 15.66$\pm$0.07    & 15.51$\pm$0.15   & 15.39$\pm$ 0.2 &104.0$\pm$13.0         &	 $-$49.0$\pm$19.0 & 13.0 $\pm$2.0   &    \nodata                          &(7);(15) \\
2MASS J2359-7335            & T6.5     & 23:59:41.09  & -73:35:04.9        &  16.17$\pm$0.04    & 16.06$\pm$0.07  &  16.05$\pm$0.13    & \nodata                                   &      \nodata                           &     12.3$\pm$1.9$^{c}$   &     \nodata   & (14) \\
%                                                 &                   &                        &                         &                                   &                                 &                                  &                                          &                                       &                           &        &                                                \\               
\tableline
\enddata
\tablecomments{(1) \citet{2007AJ....133..439C}; (2) \cite{2006AJ....131.2722C}; (3) \cite{2005AJ....130.2326T}; (4) \cite{2007ApJ...658..557B}; (5) \cite{2007MNRAS.381.1400W}; (6) \cite{2006MNRAS.373L..31M}; (7) \cite{2004AJ....127.2856B};
(8) \cite{2007AJ....134.1162L}; (9) \cite{2002ApJ...564..421B}; (10) \cite{2004AJ....127.3553K}; (11) \cite{2008MNRAS.391..320B}; (12) \cite{2000ApJ...531L..61T}; (13) \cite{2012ApJ...752...56F} ; (14) \cite{2011arXiv1108.4677K}; (15) \cite{2009AJ....137....1F};  (16) \cite{2007MNRAS.381.1400W}; (17) \citet{2007ApJ...654..570L}; (18) \cite{2004AJ....127.2948V};  (19) \cite{2003AJ....126..975T}; (20) \cite{2007A&A...466.1059K}; (21) Looper et al., in prep. (22) \cite{2012arXiv1201.2465D}; (23) \cite{2011AJ....141..203A}}\\
%\end{minipage}
\tablenotetext{a}{Optical spectral type.}
\tablenotetext{b}{$MKO$ photometry.}
\tablenotetext{c}{Distance estimated from the \cite{2012arXiv1201.2465D} absolute magnitude$-$SpT relation.}
\end{deluxetable*}

\begin{figure}
\centering
\includegraphics[width=0.5\textwidth]{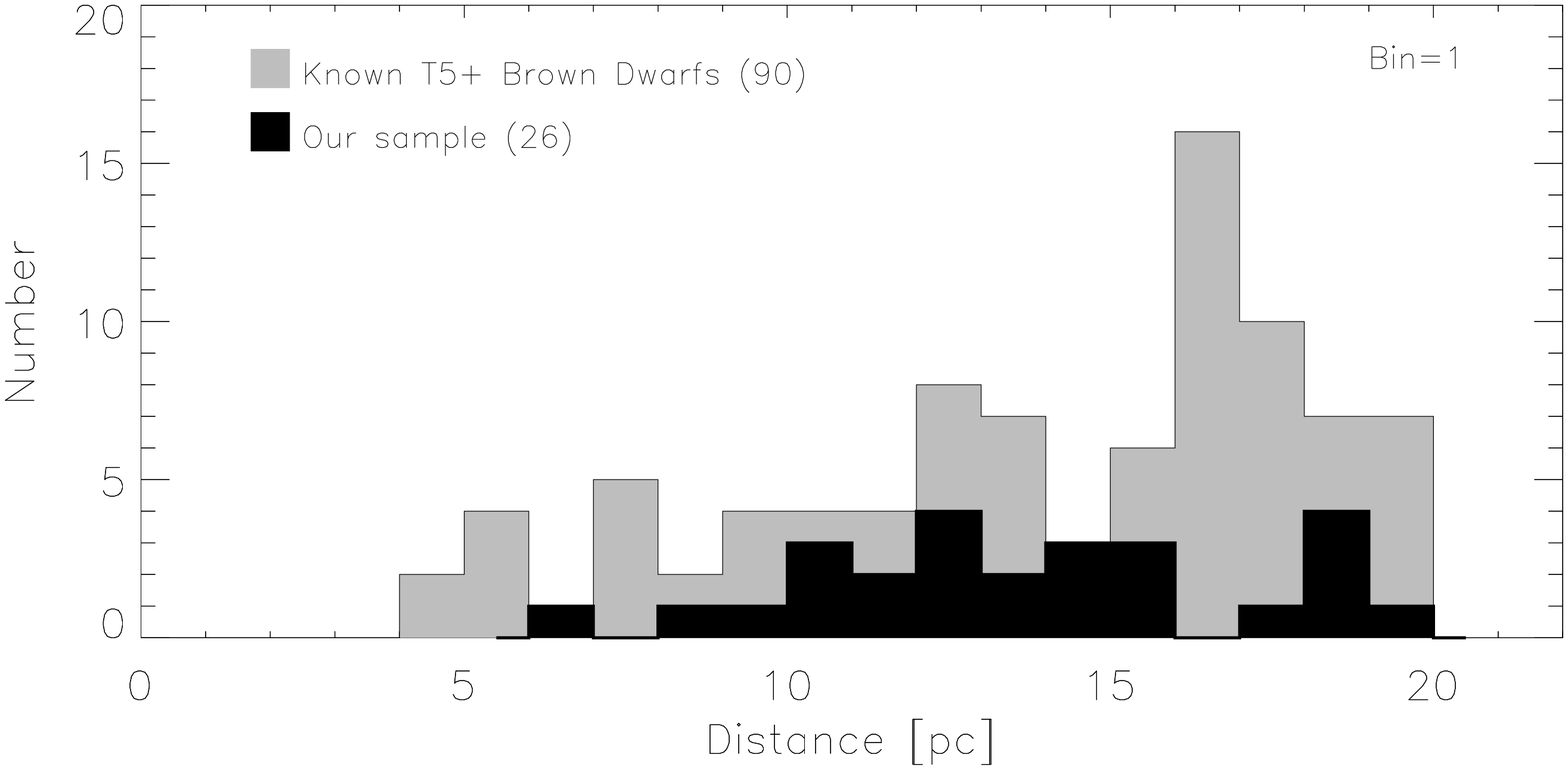}  
\caption{Number of known brown dwarfs with SpT $\geq$ T5 from Brown Dwarf Archive, \citealt{2011AJ....142...57G}, \citealt{2012AJ....144..180M}, \citealt{2013A&A...557A..43B}, \citealt{2013ApJ...764..101B} and \citealt{2014arXiv1402.1378C}. The photometric distances were determined using the \cite{2012arXiv1201.2465D} absolute magnitude$-$SpT relation.}
\label{histogram}%
\end{figure}

\begin{table*}
\scriptsize
\renewcommand\arraystretch{0.8}
\renewcommand\tabcolsep{3.0pt}
\caption{Log of observations. Magnitude limits for $\rho$=0.6$\arcsec$}             
\label{log}      
\centering          
\begin{tabular}{l ccccc c c c c c c c c}    
\hline\hline  
& \multicolumn{3}{c}{F110W} & \multicolumn{3}{c}{F127M} & \multicolumn{3}{c}{F139M} & \multicolumn{3}{c}{F164N}  \\ \cline{2-4} \cline{5-7} \cline{8-10}  \cline{11-13}
Name                             & t$^{a}$        &$\Delta$$m_{lim}$$^{b}$      & $m_{lim}$$^{c}$    &   t           &$\Delta$$m_{lim}$      &$m_{lim}$    &  t           &$\Delta$$m_{lim}$      & $m_{lim}$  & t            &$\Delta$$m_{lim}$         & $m_{lim}$  & Observation\\
                                        & (s)     &     (mag)        & (mag)            &   (s)       & (mag)            &  (mag)          & (s)         & (mag)            & (mag)          & (s)         & (mag)              &(mag)           &UT Date \\
(1)                                  & (2)     &     (3)              & (4)                  &   (5)       &   (6)               &  (7)                & (8)         &  (9)                  & (10)            & (11)       & (12)                 & (13)             & (14)  \\
\multicolumn{14}{c}{Program 11631} \\ 
\tableline                    
 2MASS J0340$-$6724          &   \nodata   &    \nodata    &    \nodata  &   413       &    2.5        &   17.25     &  413 &  2.0          &17.45    &  413    &    2.0        &    15.60            & 2011-08-06   \\ 
 SDSS J0739+6615                &   \nodata   &    \nodata     & \nodata    &   413       &     2.0        & 18.61     &   413  &  1.5          &19.82    &  413    &    2.0         & 18.03              &   2011-06-03 \\   
 2MASS J1013$-$7842          &   \nodata  &    \nodata     &  \nodata    &  413        &     2.0        & 16.19     &  413   &   2.25         &16.53     &  413   &   1.75         &    14.49           & 2011-07-23 \\ 
 2MASS J1122$-$3512          &   \nodata  &    \nodata     &  \nodata    &  413        &     1.75      & 16.45     &  413   &   1.75         &18.30    &  413    &  1.75          &    16.01            & 2011-08-13   \\  
 SDSS J1439+3042                &  \nodata   &    \nodata     &  \nodata    &  413        &     2.0        &18.78      &  413    &   1.75         &20.25   &  413    &   2.0          &18.50             &2011-07-16   \\  
 SDSS J1511+0607                &   \nodata  &    \nodata     &  \nodata    &  413        &     2.0        &17.60      &  413   &    2.0            &18.68  &  413   &     2.0        & 17.11             & 2011-05-08   \\  
 2MASS J1520$-$4422A       &   \nodata   &   \nodata     &  \nodata   &  413        &     2.25      &   16.14    &  413   & 2.0              &    16.10         &  413    &   1.5          &  14.23                    &2011-07-01   \\  

  \multicolumn{14}{c}{Program 11666}\\                 
  \tableline\tableline  
   ULAS J0034$-$0052    &   111.0      &    2.0         & 21.46   &   997   &  2.0            &19.71 &     \nodata  &   \nodata& \nodata      &   1198    &      1.75       & 21.36  &   2010-12-27   \\  
   HD3651B                        &  133.0       &    2.5         &19.94    &   997   &  1.75         &17.60  &     \nodata   &   \nodata& \nodata         &   1198    &      2.5        & 21.21   &   2010-12-30    \\
   2MASS J0050$-$3322 & 155.1       & 2.25            &19.17     &   997   &   2.25     & 17.87  &      \nodata  &   \nodata& \nodata         &   1198    &     2.25        & 18.89   &   2011-06-08   \\  
   SDSS J03255+0425    & 111.0         &  2.75            &19.92    &    997  &    2.0       & 17.78    &   \nodata    &  \nodata& \nodata        &   1198    & 2.25            & 18.75    &  2010-12-02   \\   
   2MASS J04070+1514  &  111.0       &  1.75             &18.45   &   997   &    3.0        & 18.72   &    \nodata  &    \nodata& \nodata        &   1198    &   2.25          &  18.41   & 2011-03-24   \\
   2MASS J0510$-$4208 &   67           &  2.0            &19.12   &   997    &     2.25       &  18.00 &     \nodata  &  \nodata& \nodata       &  1348     &   2.5          & 18.74  &  2011-04-03   \\   
   2MASSI J0727+1710   & 133            &   1.75           &18.13    &    997   &  2.75      & 17.93  &    \nodata   &  \nodata& \nodata        &   1198    &    2.5           &18.87           & 2010-11-11  \\  
   2MASS J0729$-$3954 &  177          &    2.0         &19.00     &   997    &    3.0        & 18.46    &    \nodata   &   \nodata& \nodata           &   1198    &    2.0         &  18.64           & 2011-05-29   \\  
   2MASS J07414+2351  &  133     &    2.0        &18.92    &   997   &      2.25        &18.23   &    \nodata     &   \nodata& \nodata           &   1198    &     1.75        & 18.29            & 2010-11-09  \\   
   2MASS J0939$-$2448 & 133      &   2.25          &19.17    &   997   &   2.0           & 17.64   &    \nodata   &   \nodata& \nodata         &   1198    &     2.0        & 18.92            &   2010-12-12   \\  
   2MASS J1007$-$4555  & 111   &   2.25          &18.64   &   997   &   2.25           & 17.85 &    \nodata   &   \nodata& \nodata          &   1198    &    1.5          &17.57            &  2011-05-14  \\  
   2MASS J1114$-$2618  & 133.     &   2.0          & 18.62  &   997   &   2.25          &17.87    &    \nodata     &  \nodata& \nodata         &   1198    &     1.5         & 18.19           &  2010-12-08 \\   
   2MASS J1231+0847      &  111     &   2.25         & 18.72    &   997     &  2.0           &17.22   &     \nodata    & \nodata& \nodata            &   1198    &    3.0         &   18.72          &   2011-07-15   \\   
   ULAS J1238+0953        & 111      &  2.25           &21.88   &   997     &   1.75           &19.98  &    \nodata     & \nodata& \nodata          &   1198    &    1.25           &21.56           & 2011-02-23   \\  
   SDSS J1250+3925       & 177      &  2.5           & 19.91   &   997    &     2.25         & 18.38 &     \nodata    &   \nodata& \nodata       &   1198    &      2.25        &18.41            &  2011-06-08  \\  
   SDSSP J1346$-$0031 & 111  &   1.75          &18.91    &   997     &  3.0           & 18.48 &    \nodata  & \nodata& \nodata            &   1198    &     2.0       &18.46              & 2011-07-17  \\
   SDSS J1504+1027       &  111       &  1.5           & 18.87   &   997     &   2.0          &18.12   &       \nodata  & \nodata& \nodata         &   1198    &     1.5         &18.88            & 2011-06-24 \\   
   SDSS J1628+2308       &  133        &    2.0         &19.29      &   997     &  2.5           &18.44  &       \nodata  & \nodata& \nodata          &   1198    &      1.25        & 18.49           & 2011-02-10  \\  
   2MASS J1754+1649    & 133         &  2.5           &18.91   &   997     &   2.0          &17.53  &       \nodata  & \nodata& \nodata         &   1198    &       1.75      &17.55             &  2011-07-19  \\ 
   SDSS J1758+4633        & 111       &  1.75            &18.65    &   997     &  2.0           & 17.70 &       \nodata  & \nodata& \nodata      &   1348    &     1.75         & 18.35        & 2010-12-09  \\
   2MASS J1828$-$4849 & 111   &  1.75           &17.95   &   997     &  1.75           &16.79   &       \nodata  & \nodata& \nodata     &   1348    &    1.75          &17.00         & 2011-10-23 \\
   2MASS J1901+4718     & 111     &   2.0           &18.45   &   997     &    2.25          &17.67 &     \nodata  & \nodata& \nodata    &   1348    &    1.5          &17.20          & 2010-12-14 \\ 
   SDSS J21241+0100     & 111     &    2.5         &19.41     &   997     &    2.25          &17.97  &     \nodata  & \nodata& \nodata    &   1198    &  2.0           &  18.19          &  2010-11-21   \\
   2MASS J2154+5942     & 177      &    2.0          &18.38    &   997     &   2.75          & 18.24 &     \nodata  & \nodata& \nodata     &   1198    &   1.75           &17.873            &   2010-12-01   \\
   2MASS J2237+7228     & 177      &   1.5          &17.90     &   1197   &    2.5         &18.10   &       \nodata  & \nodata& \nodata      &   1348    &    1.5          & 17.41         & 2011-05-18   \\   
   2MASS J2331$-$4718 & 111     &   2.25         &18.50    &   997     &  2.25           &17.58    &    \nodata  & \nodata& \nodata      &   1348    &   2.0         &17.53&2011-04-13  \\   
   2MASS J2359$-$7335 & 177   &  2.5            &19.62      &   1197  &     2.25        & 18.24  &     \nodata  & \nodata& \nodata     &      1348    &   1.75         &18.77&  2010-11-02  \\ 
  
\tableline                  
\end{tabular}
\begin{minipage}{20cm}
{$^{a}$}{Exposure time.}\\
{$^{b}$}{Limit $\Delta m$ measured from Monte Carlo simulations for each object (see Section \ref{Searchinglimits}).}\\
{$^{c}$}{Limiting magnitude calculated from Monte Carlo simulations and their respective magnitudes for each object.}\\
\end{minipage}
\end{table*}

\subsection{Imaging and Data Reduction}

\begin{figure*}
\centering
\includegraphics[width=0.7\textwidth]{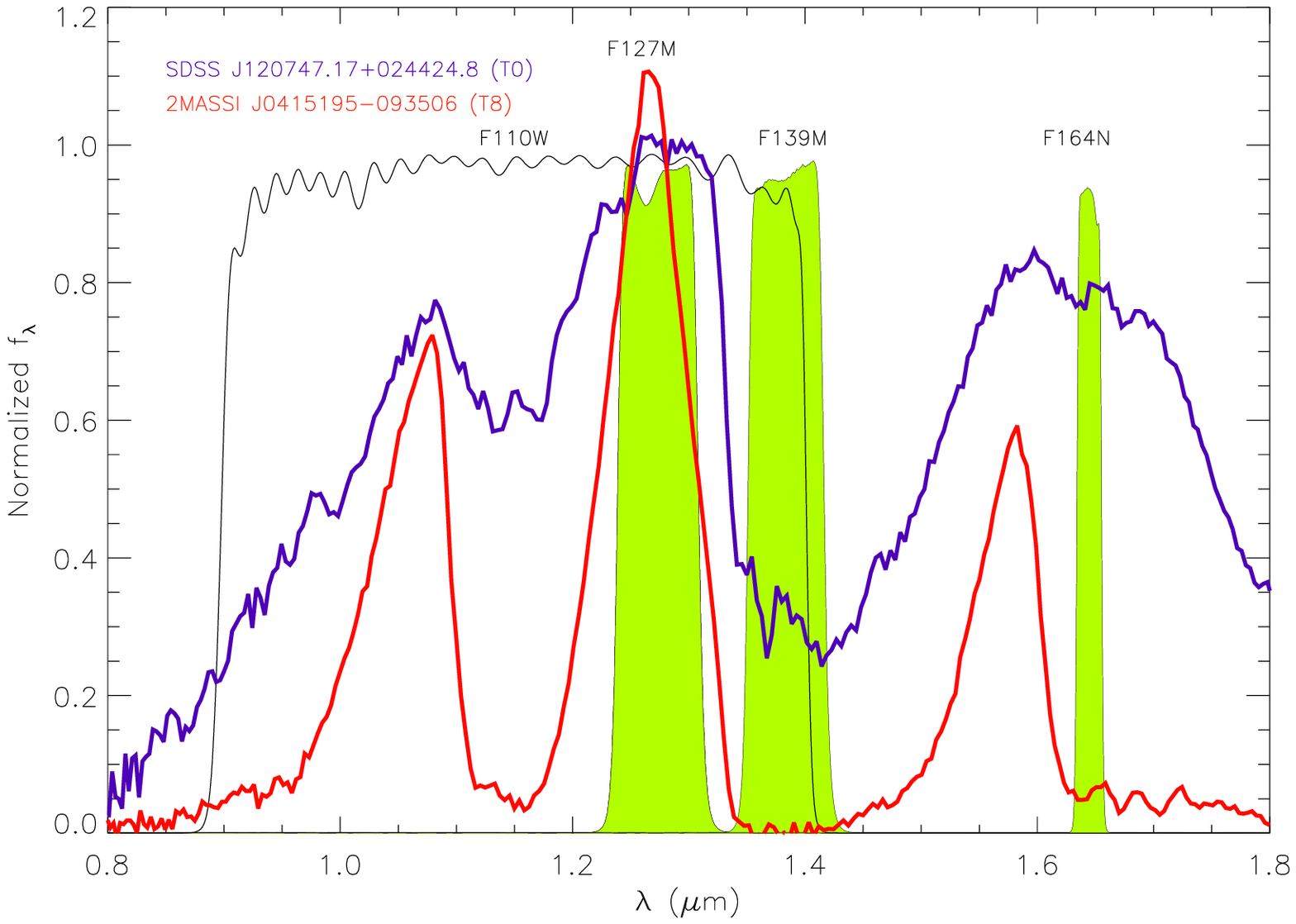}  
\caption{Filter transmission profiles of F110W  (black line), F127M, F139M and F164N (green areas), compared to the near infrared spectra of the T8  2MASS J0415-0935 \citep{2004AJ....127.2856B} and the T0 SDSS J1207+0244 \citep{2007AJ....134.1162L}}
\label{filters}%
\end{figure*}

The IR channel of WFC3 was used in both programs. The detector  (HgCdTe) is a 1024x1024 pixel array with  an angular resolution of 0.13$\arcsec$/pixel and a field of view of 123$\arcsec$x136$\arcsec$ . The camera has 16 filters covering wide (W), medium (M) and narrow (N) bands  from 800 to 1700nm. The observations analysed here use the  F110W, F127M, F139M and F164N filters (see Fig. \ref{filters}).  F110W ($\lambda_{c}$ $\equiv$ 1.1191 $\mu$m) is the widest filter covering $Y$ and $J$ bands,  encompassing the peak emission of flux from L and T dwarfs in the near infrared. F127M ($\lambda_{c}$ $\equiv$ 1.274$\mu$m) covers the 1.27$\mu$m peak  in late-T dwarfs. Finally F139M  ($\lambda_{c}$ $\equiv$ 1.3838$\mu$m) and  F164N ($\lambda_{c}$ $\equiv$ 1.6460 $\mu$m) cover $H_{2}O$  and $CH_{4}$ absorption bands in L and T dwarfs, respectively.  Program 11631 (L and early T dwarfs) utilised the F127M, F139M and F164N filters; Program 11666 (mid-late T dwarfs) data were taken using the F110W, F127M and F164N filters. \\

Images were taken in MULTIACCUM mode following a standard dither pattern (4 dithers). Exposure times ranged from 111.0s for the widest filter to 1197.7s for the narrowest (Table \ref{log}).  Due to read time limits  the images in F110W, F127M and F139M filter  (in program 11631) cover an area of 35.88\arcsec x 31.98\arcsec (276 x 246 pixels), while the images in F164N filter and F127M (in program 11666) cover 141.70$\arcsec$ x 125.32$\arcsec$ (1090 x 964 pixels). We used the pipeline processed images, which include the analog$-$to$-$digital correction, subtraction of bias and dark current, linearity correction for readout artifacts, flat$-$field image and photometric calibration. The corrected images were used as input in MultiDrizzle \citep{2002PASP..114..144F} to perform the geometric distortion correction on all individual images , cosmic-ray rejection, and the final combination of the dithered images into a single output final image. \\
 
Upon visual inspection of the images, we rejected three objects from our original sample. 2MASS J094908.6$-$154548.5 \citep{2005AJ....130.2326T} was rejected due to poor image quality. Due to its high proper motion, 2MASS J11263991-5003550 \citep{2007MNRAS.378..901F} was located outside the field of view at the time at observations. SIMP J132407.76+190627.1 \citep{2011ASPC..448..429D} was missed due to  erroneous telescope pointing. Our final sample consists of 34 sources.

\section{WFC3 Photometry}
\label{photometry}
\subsection{Measurements}

We  used  SExtractor \citep{1996A&AS..117..393B}  to measure the photometry for different  aperture sizes around sources in our final calibrated images. 
We used aperture diameters from 2 to 19 pixels (0.26\arcsec - 2.6\arcsec) around each source, and a common background annulus of 25 pixels (3.25\arcsec). The integrated counts were transformed  to Vega magnitudes with the corresponding conversion factors provided in the WFC3 instruments manual.\\ 
In order to estimate the aperture correction for our sample  we chose three isolated sources common in F110W, F127M and F164N, SDSSJ0325+0425, 2MASSJ1231+0847 and SDSSJ1346-0031. We used 2MASS J0340$-$6724, SDSS J0739+6615 and SDSS J1511+0607 for the F139M filter.  Comparison of integrated flux profiles as a function of aperture size normalized to a 20-pixel aperture demonstrates excellent agreement between the sources, with deviations of less than 0.009 mag for apertures wider than 10 pixels (see Table \ref{WFC3aperturecorrectionstable}). We adopted, a 6 pixel aperture diameter (0.78\arcsec) to extract the photometry. To calculate the total uncertainties in the magnitude corrections, we combined count uncertainties, a 1$\%$ error due to instrumental photometric stability, and 1$\%$ due to flux calibration uncertainty (see WFC3 manual\footnote{http://www.stsci.edu/hst/wfc3}). Final values are listed in Table \ref{WFC3photometry11666}.

\begin{table}[h!]
\scriptsize
\renewcommand\arraystretch{0.8}
\renewcommand\tabcolsep{0.05pt}
\caption{WFC3 Aperture Corrections.}          
\label{WFC3aperturecorrectionstable}  
\centering          
\begin{tabular}{l c c c c}     
\tableline\tableline      
 Radius   & F110W  & F127M & F39M  & F164N \\ 
(pixels)   & (mag)   & (mag)   & (mag)  & (mag)  \\
\tableline                    
 2    &  $-$1.38$\pm$0.08       &  $-$1.26$\pm$0.1        &  $-$1.08$\pm$0.05          &  $-$1.30 $\pm$0.07        \\
 3    &  $-$0.77$\pm$0.06       &  $-$0.62$\pm$0.06        &  $-$0.59$\pm$0.05         &  $-$0.72$\pm$0.04         \\
  4   &  $-$0.42$\pm$0.05       &  $-$0.36$\pm$0 .04      &    $-$0.32$\pm$0.02        &   $-$0.43$\pm$0.03         \\
  5   &  $-$0.27$\pm$0.03       &  $-$0.23$\pm$0.03        &   $-$0.21$\pm$0.01       &     $-$0.28$\pm$0.01   \\
  6   &  $-$0.20$\pm$0.03      &  $-$0.17$\pm$0.02        &    $-$0.16$\pm$0.007      &    $-$0.20$\pm$0.01      \\
  7   &  $-$0.17 $\pm$0.02      &  $-$0.15$\pm$0.02        &    $-$0.13$\pm$0.005      &    $-$0.15$\pm$0.007       \\
  8   &  $-$0.14$\pm$0.02       &  $-$0.12$\pm$0.01        &     $-$0.12$\pm$0.004      &   $-$0.13$\pm$0.006        \\
  9   &  $-$0.11 $\pm$0.01      &  $-$0.10$\pm$0.01        &    $-$0.10$\pm$0.003        &    $-$0.11$\pm$0.004      \\
  10 &  $-$0.09$\pm$0.01       & $-$0.08$\pm$0.009       &    $-$0.08$\pm$0.003       &    $-$0.09$\pm$0.003       \\
  11 &  $-$0.08 $\pm$0.01    & $-$0.07$\pm$0.008       &     $-$0.06$\pm$0.002     &    $-$0.08$\pm$0.0002       \\
  12 &   $-$0.07$\pm$0.009    & $-$0.06$\pm$0.007       &     $-$0.05$\pm$0.002     &     $-$0.06$\pm$0.001      \\
  13 &  $-$0.06$\pm$0.007     & $-$0.05$\pm$0.006       &    $-$0.04$\pm$0.003      &     $-$0.05$\pm$0.001      \\
  14 &  $-$0.05$\pm$0.006     &  $-$0.04$\pm$0.004      &     $-$0.04$\pm$0.003       &    $-$0.04$\pm$0.001       \\
  15 &  $-$0.04$\pm$0.005    & $-$0.03$\pm$0.003       &    $-$0.03$\pm$0.003        &   $-$0.03$\pm$0.001         \\
  16 &  $-$0.03$\pm$0.003    &  $-$0.02$\pm$0.002      &     $-$0.02$\pm$0.001       &   $-$0.03$\pm$0.001         \\
  17 &  $-$0.02$\pm$0.002    &  $-$0.02$\pm$0.002      &    $-$0.01$\pm$0.001        &     $-$0.02$\pm$0.0007     \\
  18 &  $-$0.012$\pm$0.001  &  $-$0.01$\pm$0.001      &    $-$0.01$\pm$0.002        &  $-$0.01$\pm$0.0009         \\
  19 &  $-$0.006$\pm$0.0006& $-$0.005$\pm$0.0006    &      $-$0.006$\pm$0.001    &    $-$0.006$\pm$0.0004       \\
\tableline                  
\end{tabular}
\end{table}

 \subsection{L and T dwarf colors}

\begin{table*}
\renewcommand\arraystretch{0.8}
\renewcommand\tabcolsep{2.0pt}
\scriptsize
\caption{WF3 Photometry for 11631 and 11666 programs.}             
\label{WFC3photometry11666}      
\centering          
\begin{tabular}{l c cc c c c}  
\tableline\tableline    
Name  &   NIR$-$SpT                     & F110W   &  F127M & F139M  & F164N  &  F127M - F164N  \\
            &   Literature                         & (mag)     & (mag)    & (mag)     &(mag)   & (mag)    \\ 
(1)       & (2)                                       & (3)           & (4)        & (5)       &    (6)            & (7)       \\
\multicolumn{7}{c}{Program 11631} \\ 
\hline                                    
2MASS J0340-6724 	& L7::         &   \nodata    &   14.76$\pm$0.02  & 15.45$\pm$0.01  & 13.60$\pm$0.01   & 1.16$\pm$0.02\\
SDSS J0739+6615 		& T1.5+/-1 &   \nodata    & 16.61$\pm$0.02  & 18.32$\pm$0.01  & 16.03$\pm$0.01   & 0.58$\pm$0.02\\
2MASS J1013-7842          & L3            &   \nodata    &  14.19$\pm$0.02  & 14.28$\pm$0.01  & 12.74$\pm$0.01  & 1.45$\pm$0.02\\ 
2MASS J1122-3512         & T2            &   \nodata    & 14.70$\pm$0.02  &16.55$\pm$0.01   & 14.26$\pm$0.01   & 0.44$\pm$0.02\\ 
SDSS J1439+3042. 		& T2.5        &   \nodata    & 16.78$\pm$0.02   & 18.50$\pm$0.01  & 16.50$\pm$0.01   & 0.28$\pm$0.02\\   
SDSS J1511+0607. 		& T2           &    \nodata    & 15.60$\pm$0.02  & 16.68$\pm$0.01  &  15.11$\pm$0.01   & 0.49$\pm$0.02\\ 
2MASS J1520-4422A       & L1.5        &    \nodata    & 13.89$\pm$0.02  & 14.10$\pm$0.01  & 12.73$\pm$0.01   & 1.16$\pm$0.02\\ 
 2MASS J1520-4422B       & L4.5       &   \nodata   & 14.57$\pm$0.02  & 15.06$\pm$0.01  & 13.60$\pm$0.01   & 0.97$\pm$0.02\\   
%                                               &                &            &                                 &                                &                                    &                       \\  
\multicolumn{7}{c}{Program 11666} \\  
\tableline\tableline    
  ULAS J0034-0052   	& T8.5 &   19.46$\pm$0.03   &  17.71$\pm$0.02   &   \nodata       & 19.61$\pm$0.02   & -1.90$\pm$0.03\\      
  HD 3651B 			& T7.5 &   16.92$\pm$0.03   &  15.98$\pm$0.02   &   \nodata       & 16.54$\pm$0.01   &-1.85$\pm$0.02\\     
  2MASS J0050-3322       & T7  &  16.92$\pm$0.03    & 15.62$\pm$0.02       &    \nodata     & 16.64$\pm$0.01    &-1.03$\pm$0.02\\                               
  SDSS J0325+0425        &  T5.5 &  17.17$\pm$0.03  &  15.78$\pm$0.02      &   \nodata     & 16.50$\pm$0.01   &-0.72$\pm$0.02\\                            
  2MASS J0407+1514      & T5 &   16.70$\pm$0.03    & 15.72$\pm$0.02       &   \nodata      &  16.16$\pm$0.01   & -0.44$\pm$0.02\\                             
  2MASS J0510-4208	&  T5 &  17.12$\pm$0.03   &  15.75$\pm$0.02      &    \nodata      & 16.24$\pm$0.01  &-0.49$\pm$0.02\\                           
  2MASSI J0727+171 	&  T7 & 16.38$\pm$0.03     & 15.18 $\pm$0.02     &    \nodata      & 16.37$\pm$0.01    &-1.18$\pm$0.02 \\     
  2MASS J0729-3954 	& T8pec &  17.00$\pm$0.03    &15.46$\pm$0.02   &    \nodata           &16.64$\pm$0.01     &-1.18$\pm$0.02\\                           
    2MASS J0741+2351      & T5 &  16.92$\pm$0.03     & 15.98$\pm$0.02    &    \nodata        &16.54$\pm$0.01     &-0.56$\pm$0.02\\                     
  2MASS J0939-2448	& T8 &  16.92$\pm$0.03     & 15.64$\pm$0.02     &    \nodata      & 16.93$\pm$0.01    &-1.28$\pm$0.02\\                    
  2MASS J1007-4555       &  T5 & 16.39$\pm$0.03     & 15.60$\pm$0.02      &    \nodata     & 16.07$\pm$0.01    &-0.47$\pm$0.02\\     
   2MASS J1114-2618 	& T7.5 &  16.62$\pm$ 0.03   & 15.62$\pm$0.02  &    \nodata    &  16.69$\pm$0.01  &-1.07$\pm$0.02\\                          
    2MASS J1231+0847    & T5.5 &  16.47$\pm$ 0.03   & 15.22$\pm$0.02   &    \nodata  & 15.73$\pm$0.01   &-0.50$\pm$0.02\\                           
  ULAS J1238+0953         &  T8.5 & 19.63$\pm$0.03    &  18.23$\pm$0.02   &    \nodata &   20.31$\pm$0.01  &-2.08$\pm$0.04\\                           
  SDSS J1250+3925         & T4 &   17.42$\pm$0.03       & 16.13$\pm$0.02   &   \nodata & 16.16$\pm$0.01   &-0.03$\pm$0.02\\                         
   SDSSP J1346-0031       &  T6.5 &  17.16$\pm$0.03   & 15.48$\pm$0.02   &   \nodata & 16.46$\pm$0.01    &-0.99$\pm$0.02\\                          
    SDSS J1504+1027       &  T7 &  17.37$\pm$ 0.03    & 16.13$\pm$0.02     &    \nodata& 17.38$\pm$0.01    &  -1.25$\pm$0.02\\     
   SDSS J1628+2308        &  T7 &  17.29$\pm$0.03     & 15.94$\pm$0.02     &   \nodata &  17.24$\pm$0.01   &  -1.29 $\pm$0.02\\     
  2MASS J1754+1649	&  T5 &  16.41$\pm$0.03   &   15.53$\pm$0.02     &    \nodata& 15.80$\pm$0.01    &   -0.27$\pm$0.02\\     
  SDSS J1758+4633         &  T6.5 & 16.90$\pm$0.03    & 15.70$\pm$0.02    &  \nodata&  16.60$\pm$0.01   &   -0.89$\pm$0.02\\     
  2MASS  J1828-4849      &  T5.5 &  16.20$\pm$0.03   &  15.04$\pm$0.02  &    \nodata &   15.24$\pm$0.01  & -0.20$\pm$0.02\\     
  2MASS J1901+4718      &  T5 & 16.45 $\pm$0.03     &  15.43$\pm$0.02    &    \nodata &   16.70$\pm$0.01   & -0.27$\pm$0.02\\     
  SDSS J2124+0100         &  T5 &  16.91$\pm$0.03     & 15.72$\pm$0.02     &    \nodata & 16.19$\pm$0.01    &   -0.47$\pm$0.02\\     
  2MASS J2154+5942      &  T5 &  16.39 $\pm$0.03    & 15.50$\pm$0.02     &   \nodata  & 16.12$\pm$0.01     &  -0.63$\pm$0.02\\     
  2MASS J2237+7228      &   T6       &  16.40$\pm$0.03  & 15.60$\pm$0.02  &    \nodata & 15.91$\pm$0.01    &  -0.31$\pm$0.02\\     
  2MASS J2331-4718       &  T5 &  16.25$\pm$0.03     & 15.33$\pm$0.02      &    \nodata & 15.53$\pm$0.01    &   -0.19$\pm$0.02\\     
  2MASS J2359-7335       &  T6.5      &  17.12 $\pm$0.03   & 16.00$\pm$0.02   &    \nodata& 17.02$\pm$0.01    &  -1.03$\pm$0.02\\         
\tableline                  
\end{tabular}
\end{table*}

\begin{figure*}[h!]
\centering
\includegraphics[width=0.7\textwidth]{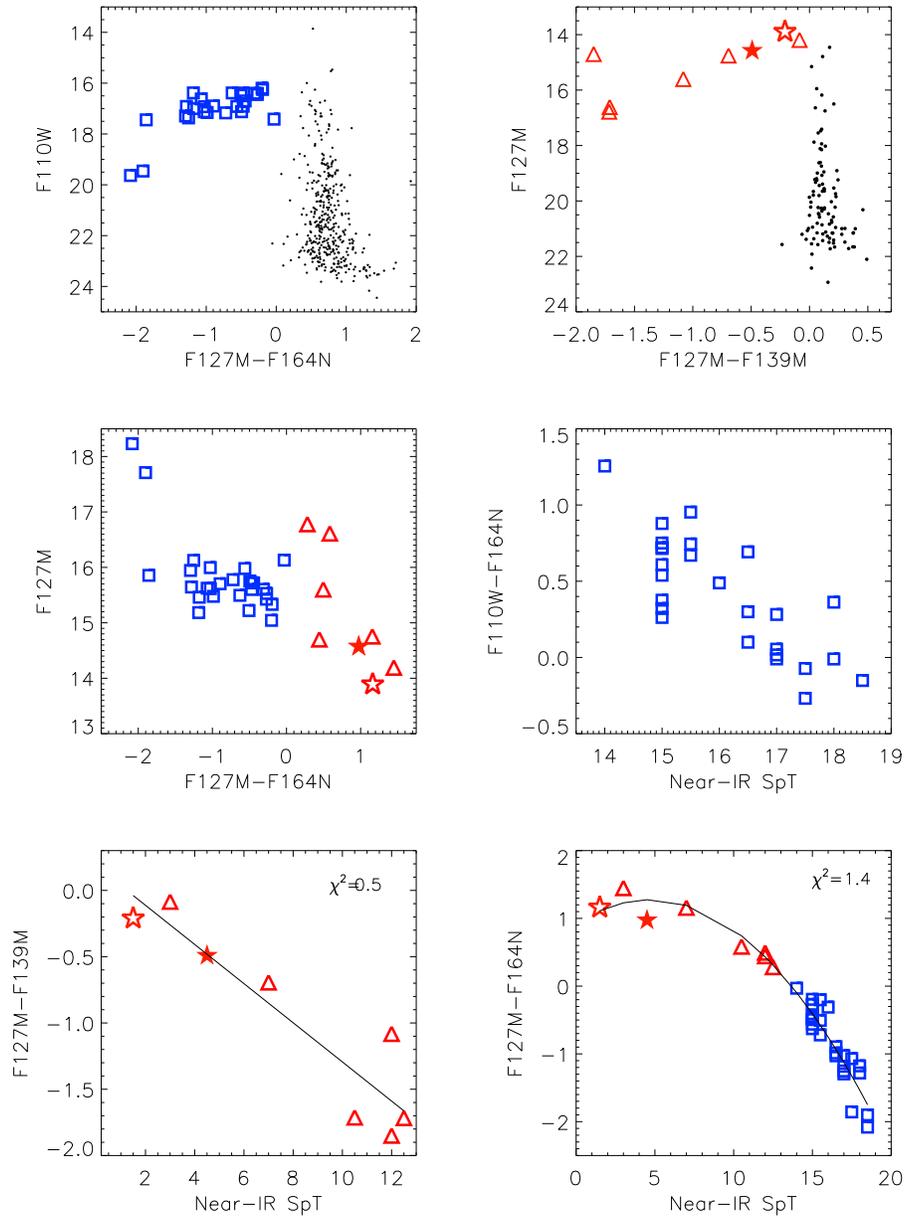} 
\caption{Segregation of L and T dwarfs with WFC3 photometry. Magnitude vs. color and color vs. NIR SpT  for 11666 (full blue circles) and 11631 (red triangles) programs, including the known binary system  2MASS J1520$-$4422AB (red stars), respectively. Background sources are represented by black points. Linear (for F127M-F139M) and quadratic (for F127M-F164N) fits to the photometric data are indicated by the solid lines. Spectral types are encoded as SpT(L0) = 0, SpT(T5) = 15 and, SpT(Y0) = 20. The uncertainties are smaller than the symbol size.}
\label{bothfigures}%
\end{figure*}

To provide adequate color discrimination of L and T dwarfs from background sources, we examined all possible color combinations. While the majority of background sources have neutral NIR colors, our targets show very red colors (see Table \ref{WFC3photometry11666}) due to molecular absorption. Figure \ref{bothfigures} displays the magnitude vs. color and color vs. NIR-SpT\footnote{Except for 2MASSI J0340$-$6724, 2MASS J1013$-$7842 and 2MASS J2237+7228 where only the optical spectral type was available.} of our targets.  F110W$-$F164N shows considerable scatter vs. spectral type, so we did not calculate a spectral type relation for this color. This is likely due to the width at the F110W filter.  F127M - F139M color displays a strong trend with spectral type in the late$-$L and early$-$T dwarfs. A linear fit to F127M$-$F139M color yields,

\begin{equation}
\small
SpT = 1.56 - 6.25*(F127M-F139M)\\
\end{equation}

where SpT(L0) = 0, SpT(T5) = 15 and, SpT(Y0) = 20. The scatter is 1.2 subtypes.\\

Similarly, a quadratic fit of F127M$-$F164N color to SpT yields,
\begin{equation}
\small
SpT=13.22 - 5.37*(F127M-F164N) -1.56*(F127M-F164N)^2
\end{equation}
The scatter in this relation is 0.6 subtypes and hence this color is a more accurate proxy for spectral type than F127M-F139M color. \\

Both trends reflect the strengthened $H_{2}O$ and $CH_{4}$ absorption bands along the L and T sequence. Because these bands saturate, continuum  fluxes also decline in the Y dwarf regime (\citealt{2011arXiv1108.4678C}; \citealt {2011arXiv1108.4677K}), so the trends may not persist to arbitrarily low temperatures.

\section{WFC3/HST PSF FITTING ANALYSIS}
\label{psffitting}
\subsection{Method}
The main goal of this study is to identify binary systems in the sample. There are no well-resolved pairs  other than the previously identified 2MASS J1520$-$4422AB (\citealt{2007ApJ...658..557B}; \citealt{2007MNRAS.374..445K}). To identify more closely-separated pairs with blended PSFs, we used  a PSF-fitting algorithm similar to that described in \cite{2003ApJ...586..512B}. \\

In previous studies (\citealt{2006ApJS..166..585B}; \citealt{2012arXiv1201.2465D}), the Tiny Tim\footnote{http://tinytim.stsci.edu/cgi-bin/tinytimweb.cgi} program \citep{1995ASPC...77..349K} has been used to generate a super$-$sampled PSF model that takes into account the source SED and instrument response.  This tool does not currently implement oversampling for WFC3, so we decided to generate PSF models of each filter from the data. The WFC3 diffraction limit goes from 0.096$\arcsec$ for the bluest filter, F110W, to 0.142$\arcsec$ for the reddest one F164N. Therefore, we extracted subimages of 20x20 pixels (2.6\arcsec x 2.6\arcsec) centered on point sources, which were then resampled at ten times the original pixel size and recentered by subpixel offsets. We median-combined 83 background sources in F110W, 399 sources in F127M, 12 sources in F139M and 515 sources in F164N images to create the PSF models. We found these models provide superior fits to the target  than the WFC3 TinyTim model.\\

To search for faint companions, we used an iterative routine focused on the same 2.6\arcsec x 2.6\arcsec subimages centered on each target. First, initial guesses for the positions and fluxes for two components were made using a simple peak detection on the original image (for the primary)  and on the residual image after PSF subtraction (for the secondary). 
The routine then finds  the optimal primary and secondary position by shifting in steps of 0.1 pixels  and  flux scaling in steps of 1$\%$ (0.01mag). 
The quality of fit was assessed  using the $\chi^2$ statistic:

\begin{equation}
\small
\chi^2 = \sum_{ij} \frac{(D_{ij}-\alpha M_{ij})^2}{\sigma_{ij}^2}\\
\end{equation}

\begin{figure*}
\centering
\includegraphics[width=0.6\textwidth]{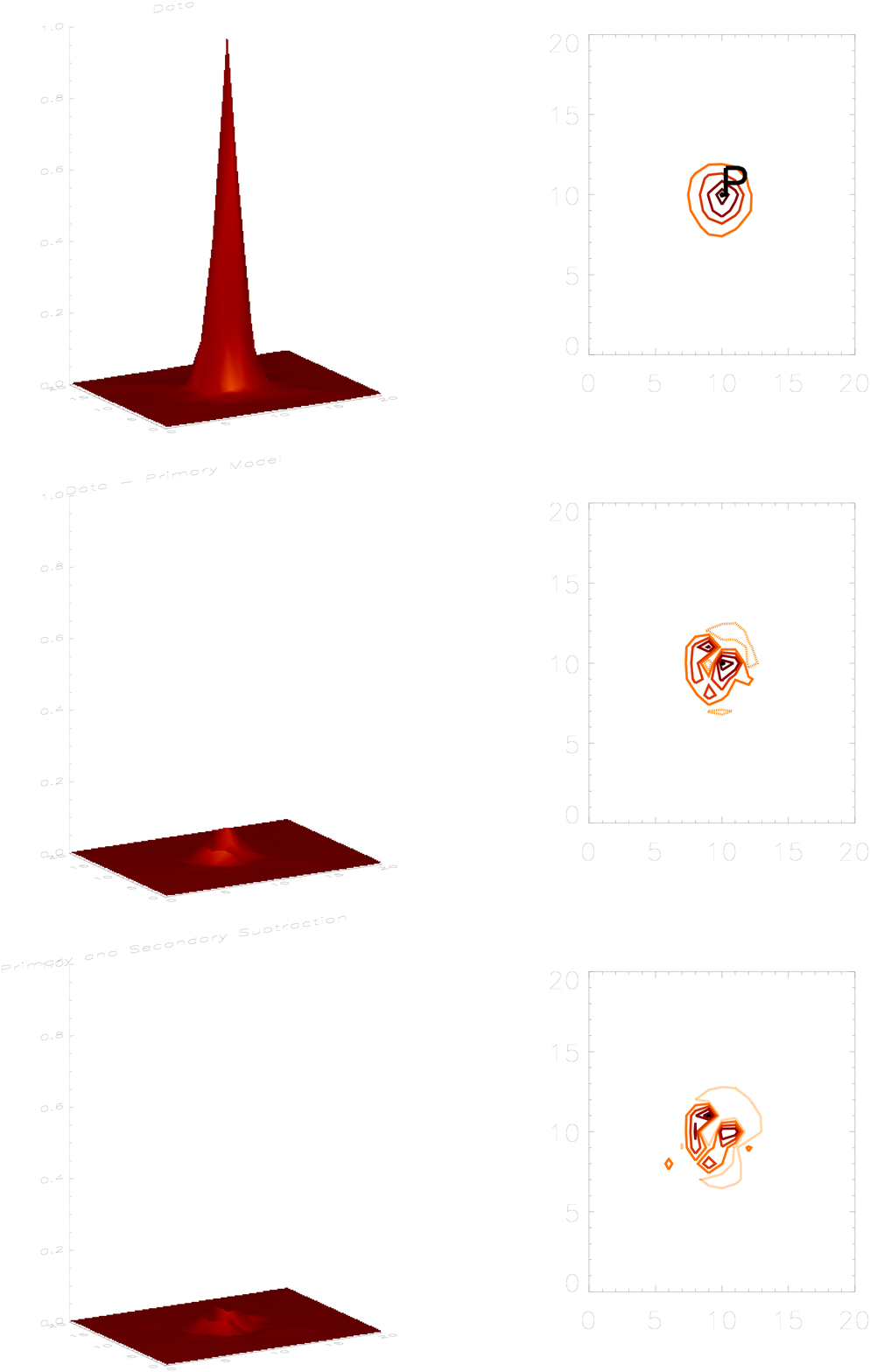}
\caption{2MASS J0727+1710 in F127M filter. In the upper part of the figure are shown the surface and contour plots previous to the PSF subtraction. The red letter 'P' represents where  the primary BD's coordinates are located. In the middle and bottom part of the figure are shown the surface and contour plots for the residuals after the primary PSF-model subtraction and the residuals after primary and secondary PSF-models subtraction, respectively. The contour levels represent the --0.3,  --0.2, --0.1 (dashed lines), 0.1, 0.3, 0.5, 0.8, 0.95 (solid lines) of the maximum flux from the data, and after the primary and secondary PSF-subtraction.}
\label{ib1n08020}
\end{figure*}

Here $D_{ij}$ is the data, $M_{ij}$ is the model, $\alpha$ is the scaling value between the data and model and $\sigma_{ij}^2$ is the data variance.  
An illustration of this algorithm  is shown in Figure \ref{ib1n08020}. Fits were done with both single and binary PSF models, and to assess the statistical significance of the latter we use the one-side F-test.\\

\begin{equation}
\small
F= \frac{min (\chi^{2}_{single})/ \nu_{s}}{min (\chi^{2}_{binary})/\nu_{b}}\\
\end{equation}

where $\nu_{s}$ and $\nu_{b}$ are the degrees of freedom for the singles and binary fits, respectively, because some parts of the image do not contribute to the fit (i.e. regions with no source flux). So, the degrees of freedom were calculated taking into account the effective pixels involved in the fitting,
\begin{equation}
\small
Pixels_{eff} = \frac{\sum_{ij} M_{ij}}{max (M_{ij})}\\
\end{equation}

\begin{equation}
\small
\nu_{s,b} = Pixels_{eff}- N_{parameters}
\end{equation}

where $N_{parameters}$ is 3 for the single model, and 6 for the binary (\citealt{2010ApJ...710.1142B}).

\begin{table}
\renewcommand\arraystretch{0.5}
\renewcommand\tabcolsep{1.2pt}
\scriptsize
\caption{The statistical Analysis for F127M filter.}             
\label{F_testF127M}      
\centering          
 \begin{tabular}{l l  cc c c cc c}
\tableline\tableline    
Name           & $\chi^2_{single}$         & $\chi^2_{binary}$ &  F-test  &   $\rho^{a}$ &   $\rho^{a}$           & P.A.$^{b}$      &$\Delta m$ \\
                     &                                  &                         & ($\%$) &   (pixel)  &  ($\arcsec$) & $\fdg$ & (mag)\\
                     (1)                                &  (2)                 &  (3)        &   (4)       & (5)                     & (6)      & (7)    &(8)  \\     
\multicolumn{7}{c}{Program 11666}\\
\tableline
2MASS J0340$-$6724    & 95.9      & 99.4  & 44  & 0.83   & 0.108 &  69.53& 2.82\\
SDSS J0739+6615          & 288.7    & 328.0  & 40  & 0.81   &  0.106 & 184.96 &2.08\\
2MASS J1013$-$7842    & 70.7      & 52.4   & 60  & 1.51      & 0.197 & 253.44 &2.18\\
2MASSJ1122$-$3512     & 5328.8 & 4908.3  & 43  & 0.06    & 0.008 & 191.42 & 2.11\\
SDSS J1439+3042          & 100.2    & 115.7    & 41  & 1.19      & 0.155 & 118.42&2.14\\
SDSS J1511+0607          & 270.7    & 282.3    & 43   & 0.55      & 0.071 & 233.01 &2.14\\
2MASS J1520$-$4422A & 0.1         & 0.01    & 100 & 9.23       & 1.200 & 29.65 &0.04\\
&   && &  & & & \\
\multicolumn{7}{c}{Program 11666}\\
\tableline
ULAS J0034$-$0052     & 42.2    & 42.1   &  45  & 0.61     & 0.080 &68.11      &2.31 \\
HD 3651B                       & 1878.6  & 1721.7 &  44  & 0.08 & 0.011 & 	21.14	   &2.16\\
2MASS J0050$-$3322  & 400.0 & 369.7  &  44  & 0.12     & 0.015 & 259.21 	  &2.43\\
SDSS J0325+0425         & 686.9 & 867.8  &  36   & 0.88    & 0.115 &  115.58	  &2.27\\
2MASS J0407+1514     & 16.3    & 14.8    &  50   & 0.83    & 0.108 & 27.53	   &3.20\\
2MASS J0510$-$4208 & 149.1  & 162.3 &  43    & 0.97 & 0.126 &  20.59	  &2.35\\
2MASSI J0727+1710    & 173.8  & 166.0  & 43     & 0.0    & 0.0    &  	 304.01 	&3.08\\
2MASS J0729$-$3954  & 93.3    & 84.7     & 48     & 1.07 & 0.139 &  98.61	  &3.08\\
2MASS J0741+2351      & 77.1   & 62.9     & 52  & 1.75     & 0.227 &  51.73	  &2.55\\
2MASS J0939$-$2448  & 68.1    & 49.5     & 59   & 1.84 & 0.239 & 122.55	   &2.39\\
2MASS J1007$-$4555  & 58.7    & 41.7     & 60   & 1.60 & 0.208 &30.35 	  &2.38\\
2MASS J1114$-$2618  & 307.7  & 302.6   & 41    & 0.14 & 0.018 &136.22 	    &2.93\\
2MASS J1231+0847      & 513.6  & 473.0   & 44    & 0.09 & 0.011 & 69.78	   &2.30\\
ULAS J1238+0953         & 56.5   & 79.2    & 33  & 0.74     & 0.096 & 312.23	   &2.02\\
SDSS J1250+3925        & 33.4      & 34.3        & 45    & 0.62 & 0.080 & 143.86	   &2.58\\
SDSSP J1346$-$0031 & 453.4  & 413.1    & 44    & 0.13 & 0.017 &219.76	    &3.25\\
SDSS J1504+1027        & 1100.8 & 1358.7  & 37     & 0.88& 0.114 &  82.03 	  &2.18\\
SDSS J1628+2308        & 1139.3 & 1070.8    & 43     & 0.06 & 0.008 & 	24.22    &2.74\\
2MASS J1754+1649     & 44.2      & 53.7      & 38     & 1.34 & 0.174 &  87.38 	 &2.28\\
SDSS J1758+4633        & 571.6   & 705.5        & 37      & 0.88 & 0.114 & 122.45	   &2.36\\
 2MASS J1828$-$4849 & 99.6    & 58.5        & 68     & 1.21& 0.157 & 104.86	   &1.89\\
2MASS J1901+4718      & 1203.8 & 1132.1   & 42      & 0.0 & 0.0     & 201.57	   &2.39\\
 SDSS J2124+0100        & 247.0   & 234.8      & 42      & 0.06 & 0.008 & 154.92 	 &2.65\\
 2MASS J21547+5942     & 472.6  & 448.1    & 42   & 0.12 & 0.0161 & 63.44	  &2.59\\
2MASS J2237$-$7228  & 116.3  & 108.5     & 45     & 1.06 & 0.138 & 307.22	   &2.99\\
 2MASS J2331$-$4718 & 106.0  & 54.5       & 69    & 1.59 & 0.206 &  44.28	  &2.39\\
2MASS J2359$-$7335  & 116.2  & 94.4     & 53    & 1.82& 0.236 & 15.22	  &2.53\\
\tableline      
\tableline                  
\end{tabular}
\begin{minipage}{9cm}
{$^a$}{Separation between the primary and secondary components after the PSF fitting.}\\
{$^b$}{Position angle.}\\
\end{minipage}
\end{table}

\subsection{Results}
The results of these fits  are summarized in Table \ref{F_testF127M} for  F127M. The only binary system identified by our PSF$-$fitting routine was the previously-known wide binary, 2MASS J1520$-$4422AB with separation 1.20$\pm$0.01\arcsec and PA=29$\fdg$65$\pm$0$\fdg$70. These values are marginally consistent with previous determinations {\citep{2007ApJ...658..557B}}. We can conclude that our routine gives us reliable results for resolved BD binary systems. \\

No other sources were found to be significally better fit by a binary PSF model, implying that they are single or unresolved with WFC3's resolution and sensitivity.\\

\begin{figure*}
\centering
\includegraphics[width=0.25\textwidth]{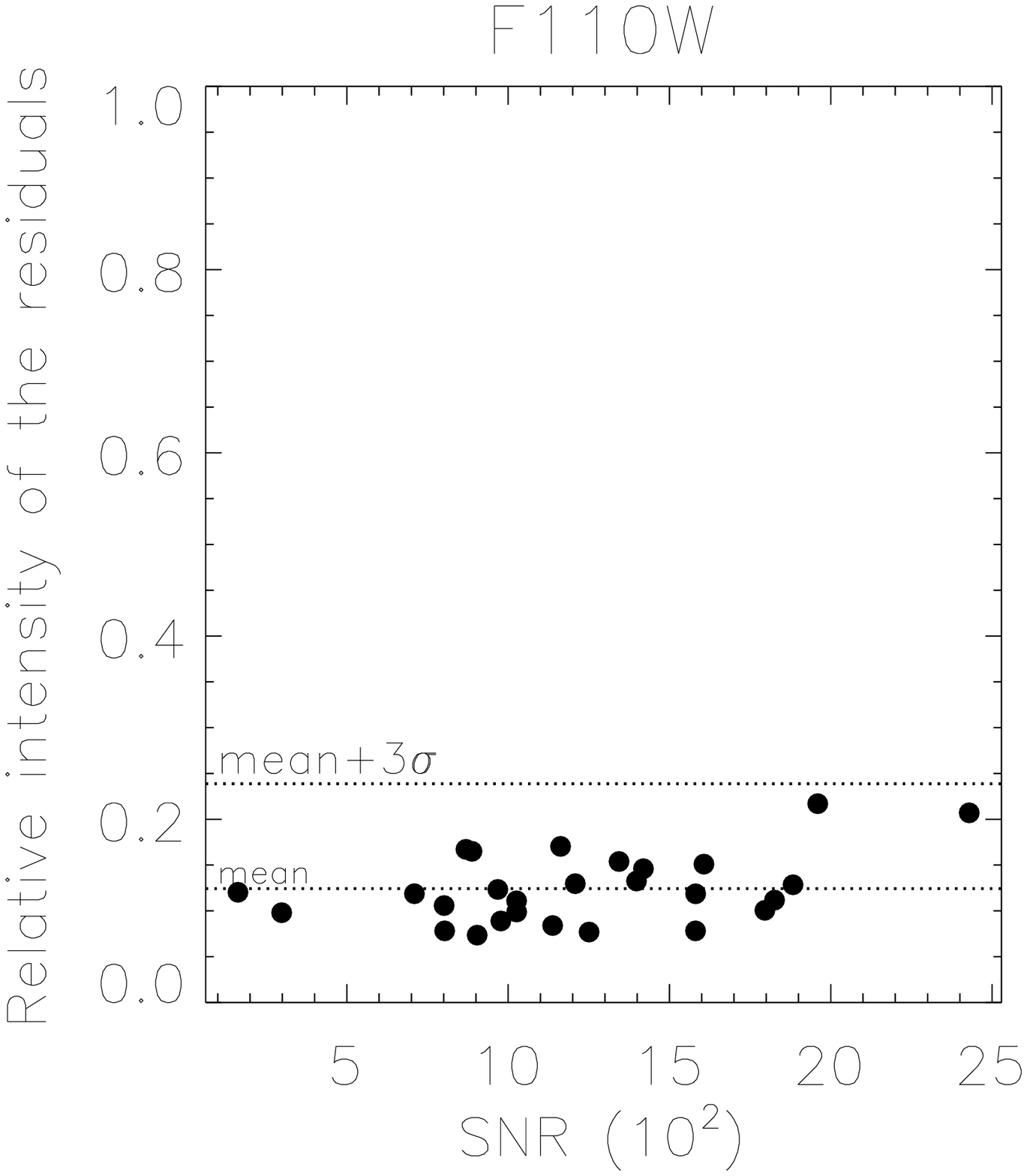}\includegraphics[width=0.25\textwidth]{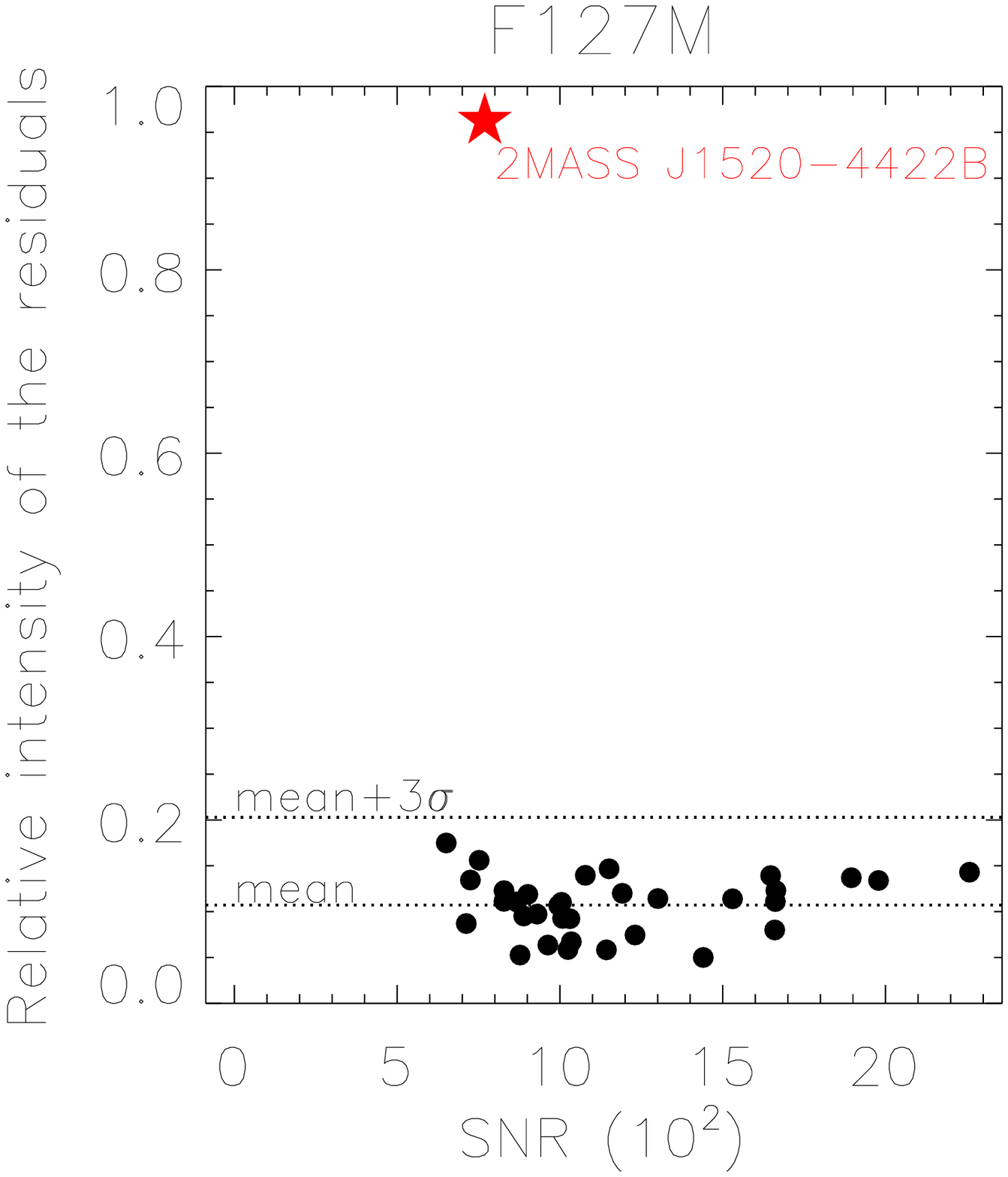}\includegraphics[width=0.25\textwidth]{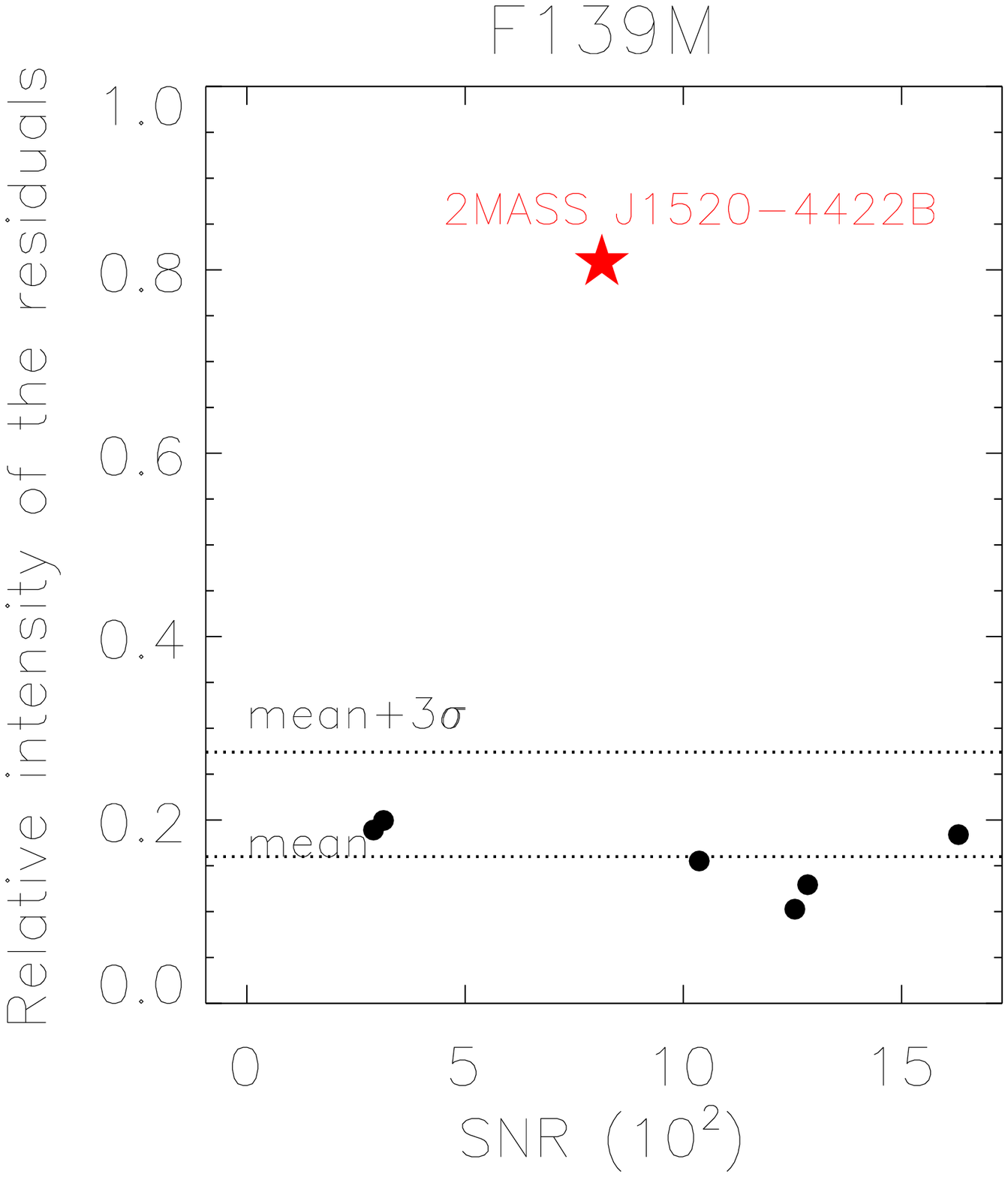}\includegraphics[width=0.25\textwidth]{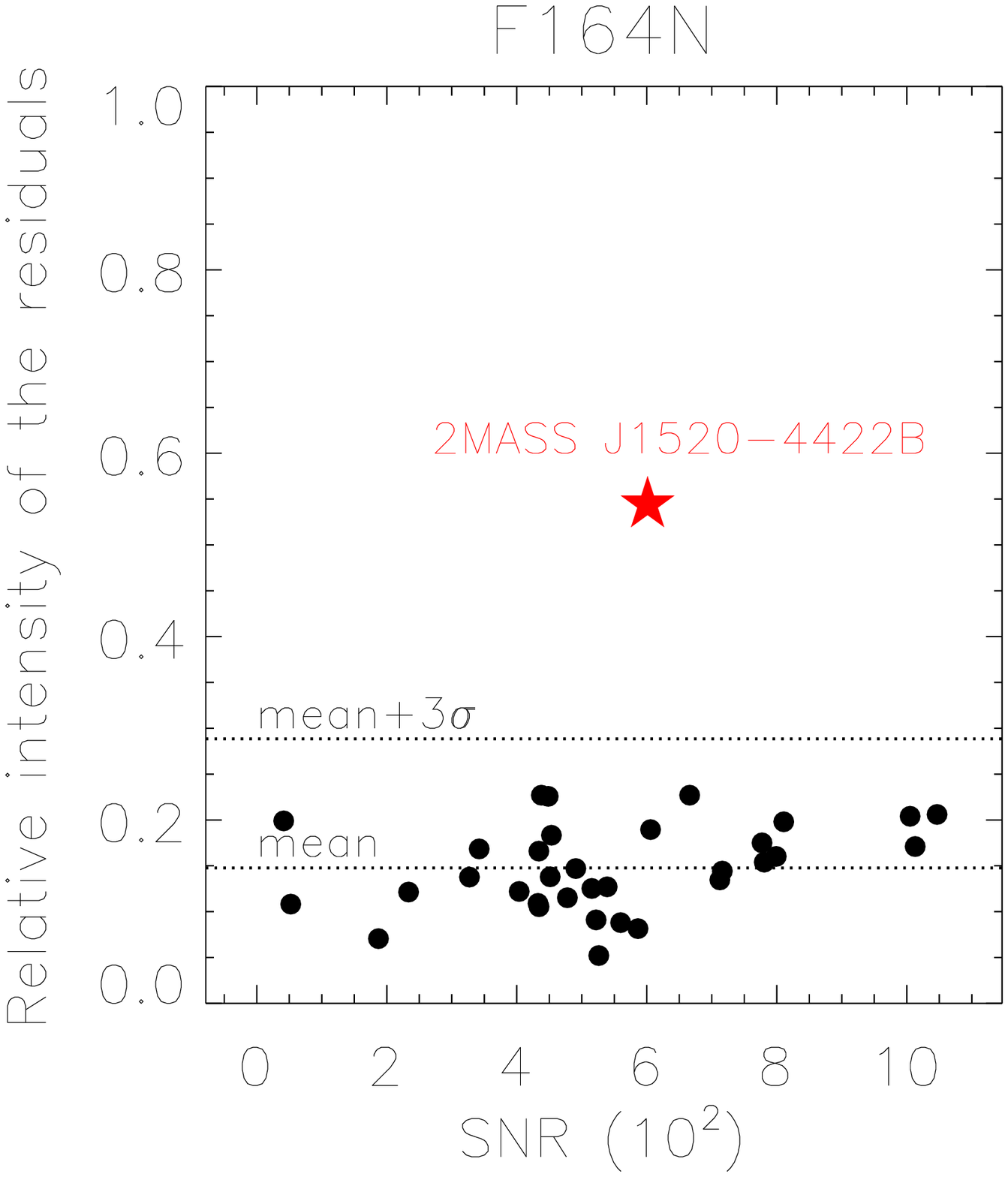}\\
\caption{Relative intensity of the residuals after the primary PSF subtraction as a function of the Signal to Noise Ratio. The only object detected is 2MASS J1520$-$4422B, which is shown with red star above the mean + 3$\sigma$ value.}
\label{residuals_vs_SNR}
\end{figure*}

To check our results is to quantify the relative intensity of the residuals after the primary PSF subtraction.  Figure \ref{residuals_vs_SNR} shows the result of this analysis showing images after the PSF subtraction.  The 2MASS J1520$-$4422AB system clearly has the highest residuals compared to the rest of the sample because of its resolved secondary. None of the others targets show clear evidence of multiplicity.

\subsection{Searching limits}
\label{Searchinglimits} 
To assess our detection limits for mid-late T companions we performed a multi-step Monte Carlo simulation to calculate the detection and false positive rates as a function of separation and relative magnitude for each source in three WFC3 filters (F110W, F127M and F164N).  Our simulation used $10^{5}$ fake stars (generated from the PSF model) implanted around each target with different orientations, distances (from 1 to 6 pixels)  and $\Delta m$ (from 0 to 5 mag).  Our PSF-fitting routine was then used to recover the implants with steps in distance and magnitude of 0.5 pixels  and 0.2 magnitudes to find the limit beyond which the fake stars are not correctly recovered. To quantify the effect of false positives, we performed another Monte Carlo simulation adding $10^{3}$ variations  of Gaussian noise to each image and  seeing where a (false) secondary is found.\\

\begin{figure*}
\centering
\includegraphics[width=0.30\textwidth]{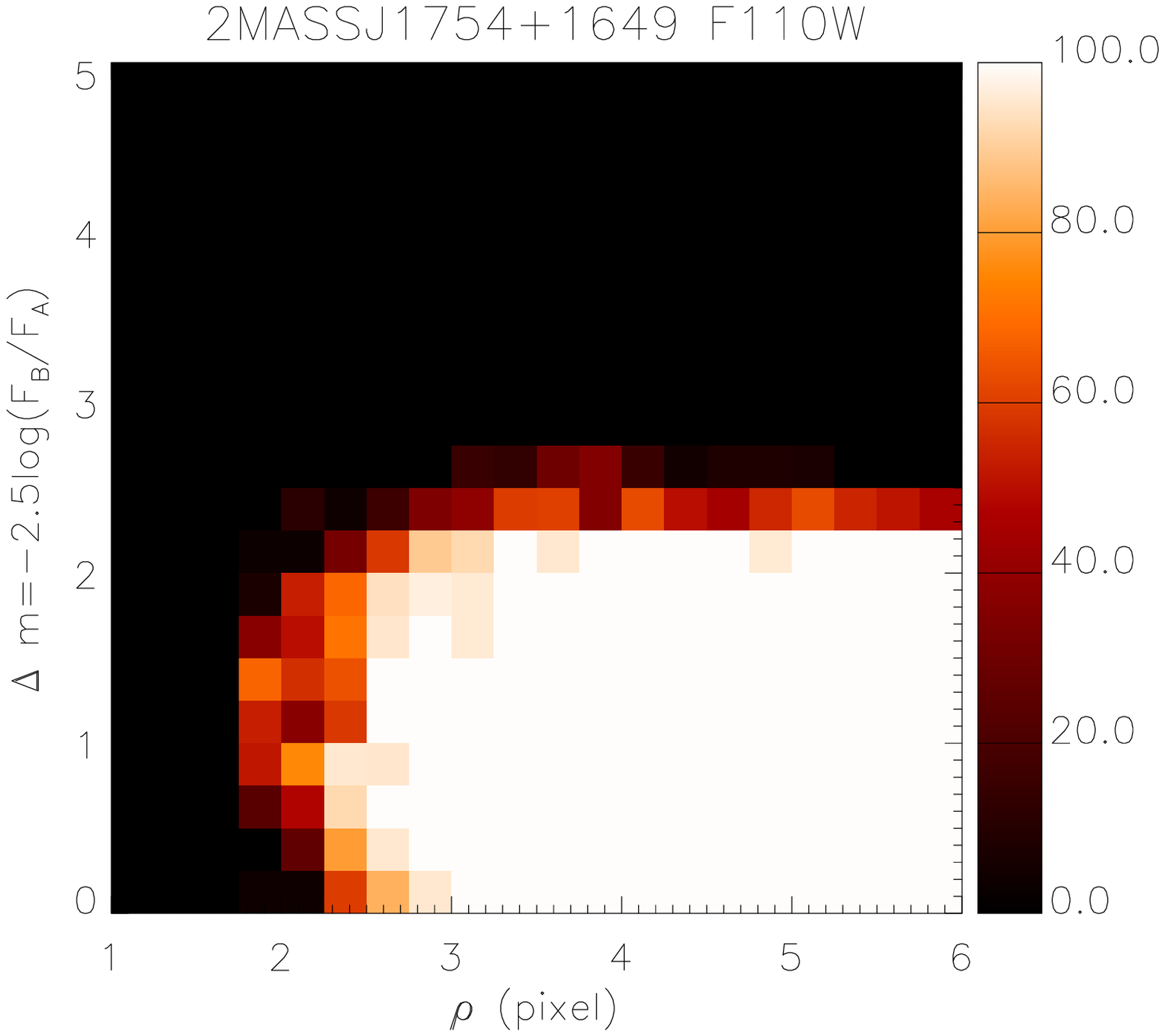}\includegraphics[width=0.30\textwidth]{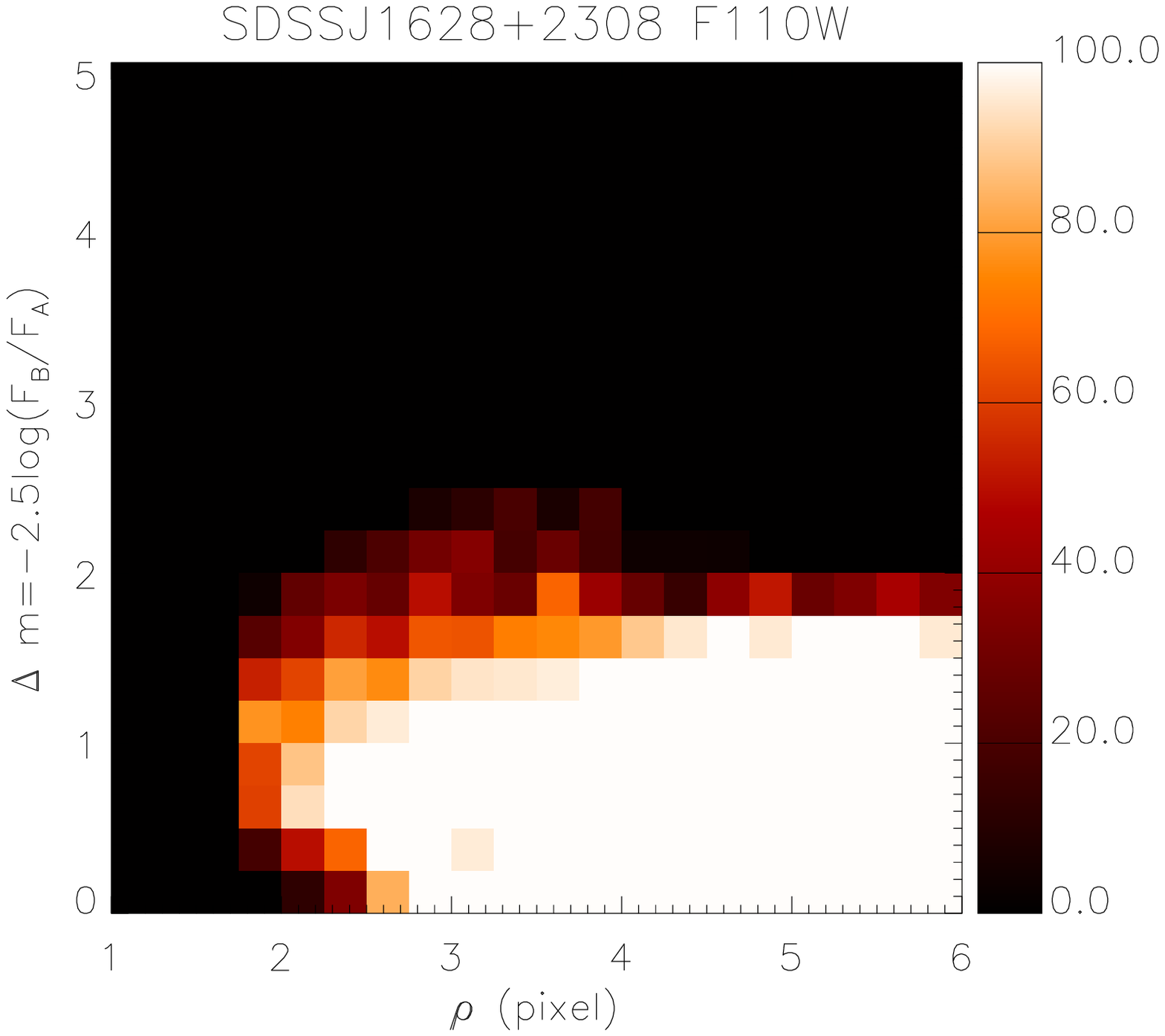}\\
\includegraphics[width=0.30\textwidth]{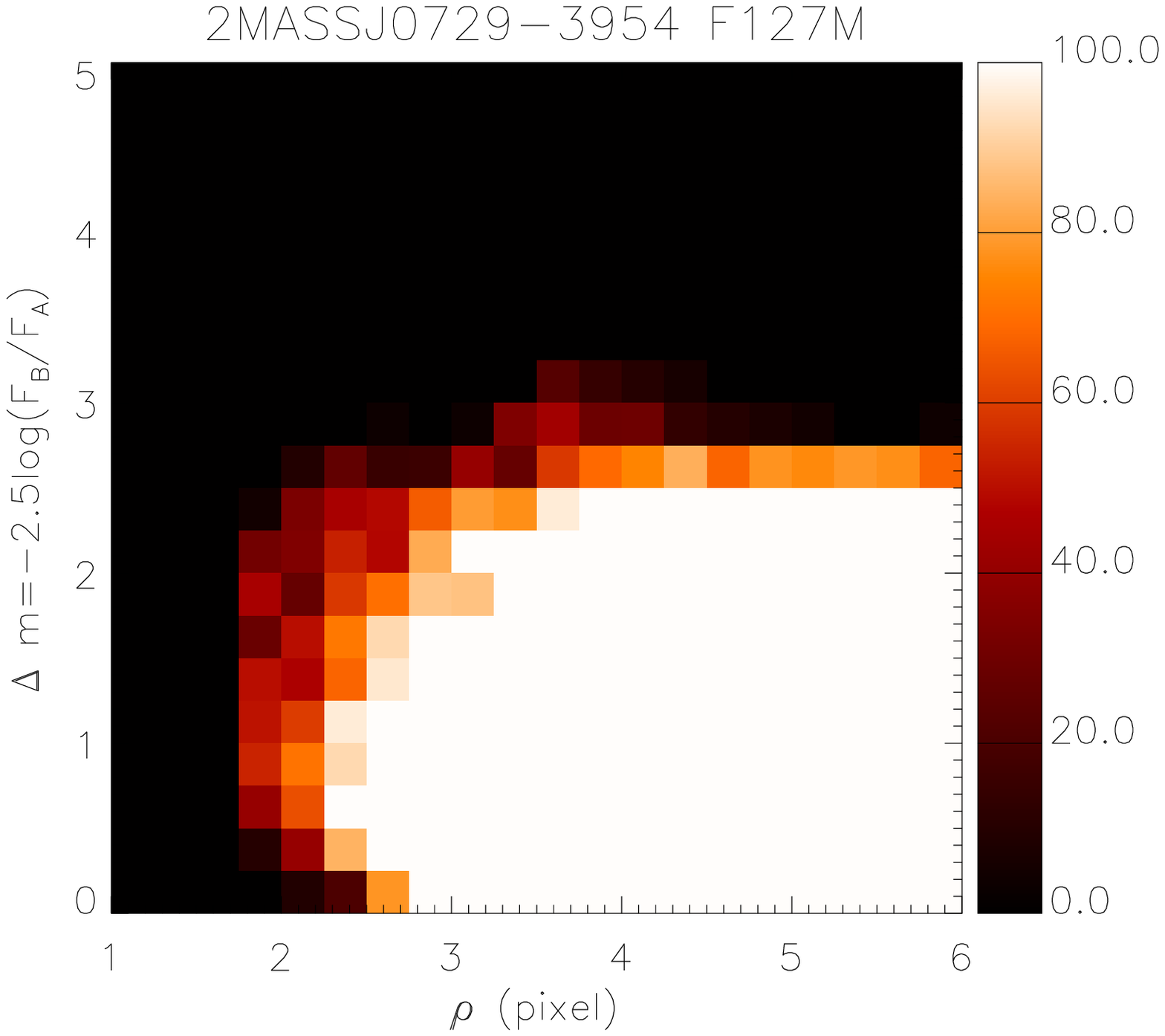}\includegraphics[width=0.30\textwidth]{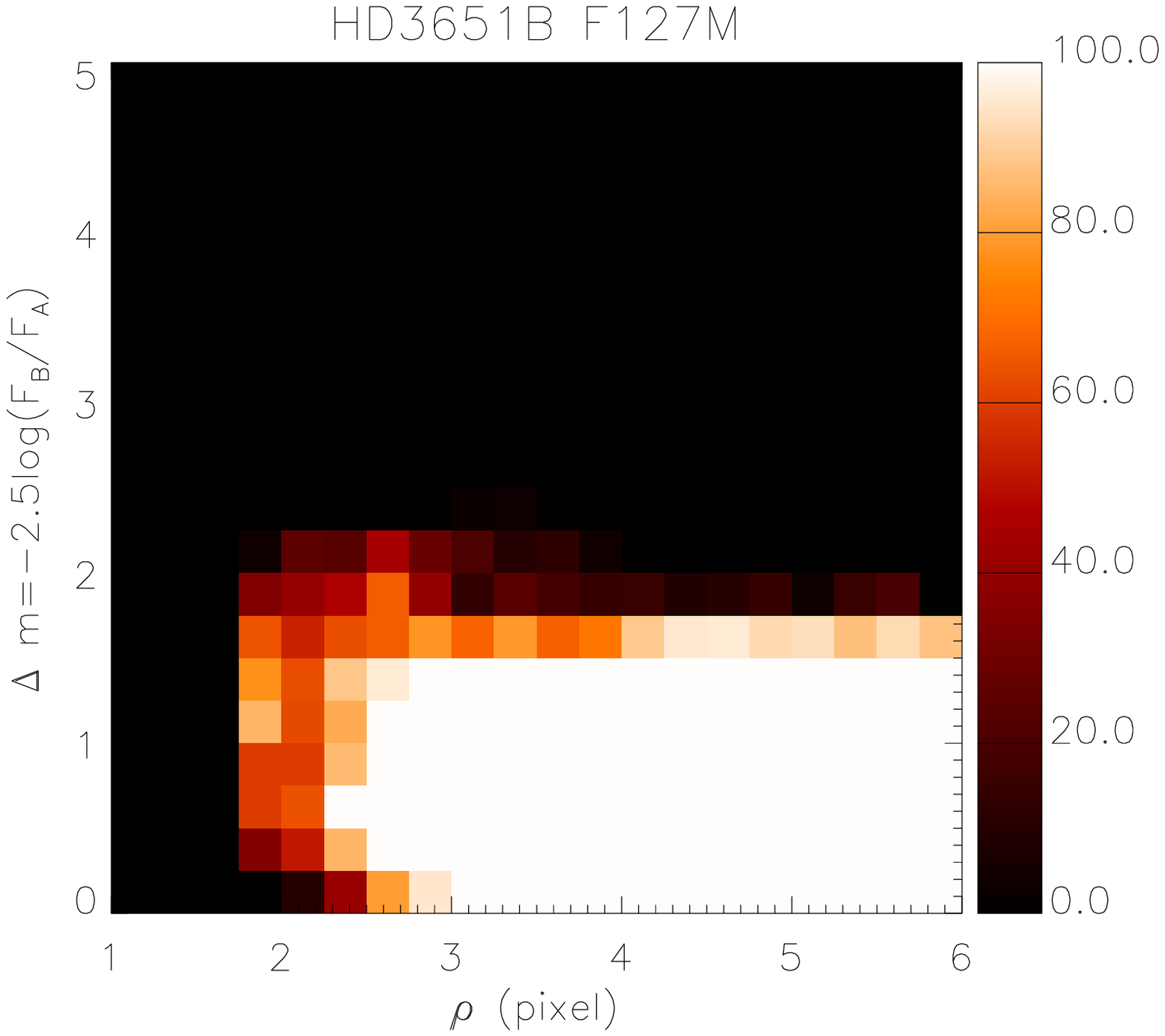}\\
\includegraphics[width=0.30\textwidth]{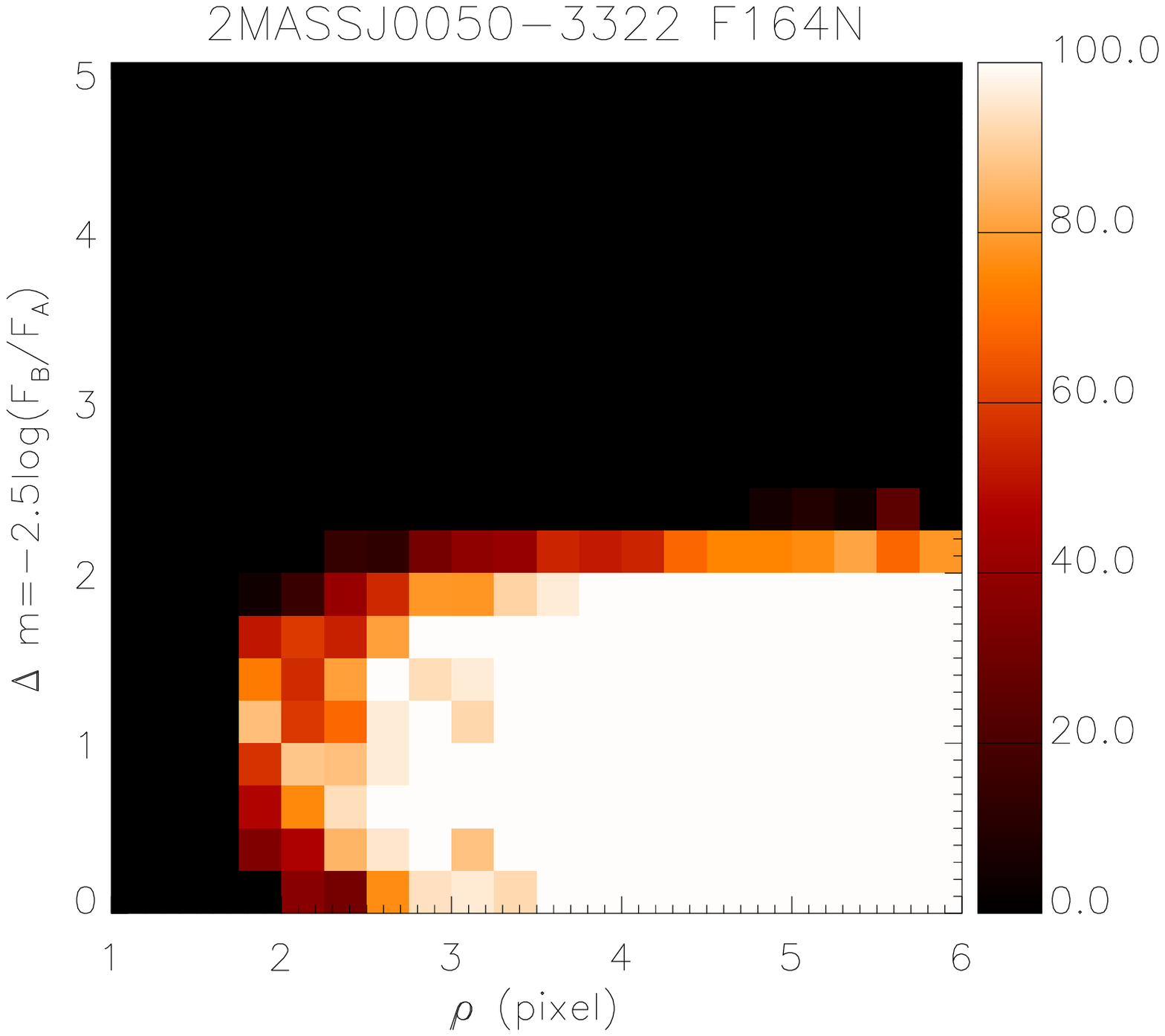}\includegraphics[width=0.30\textwidth]{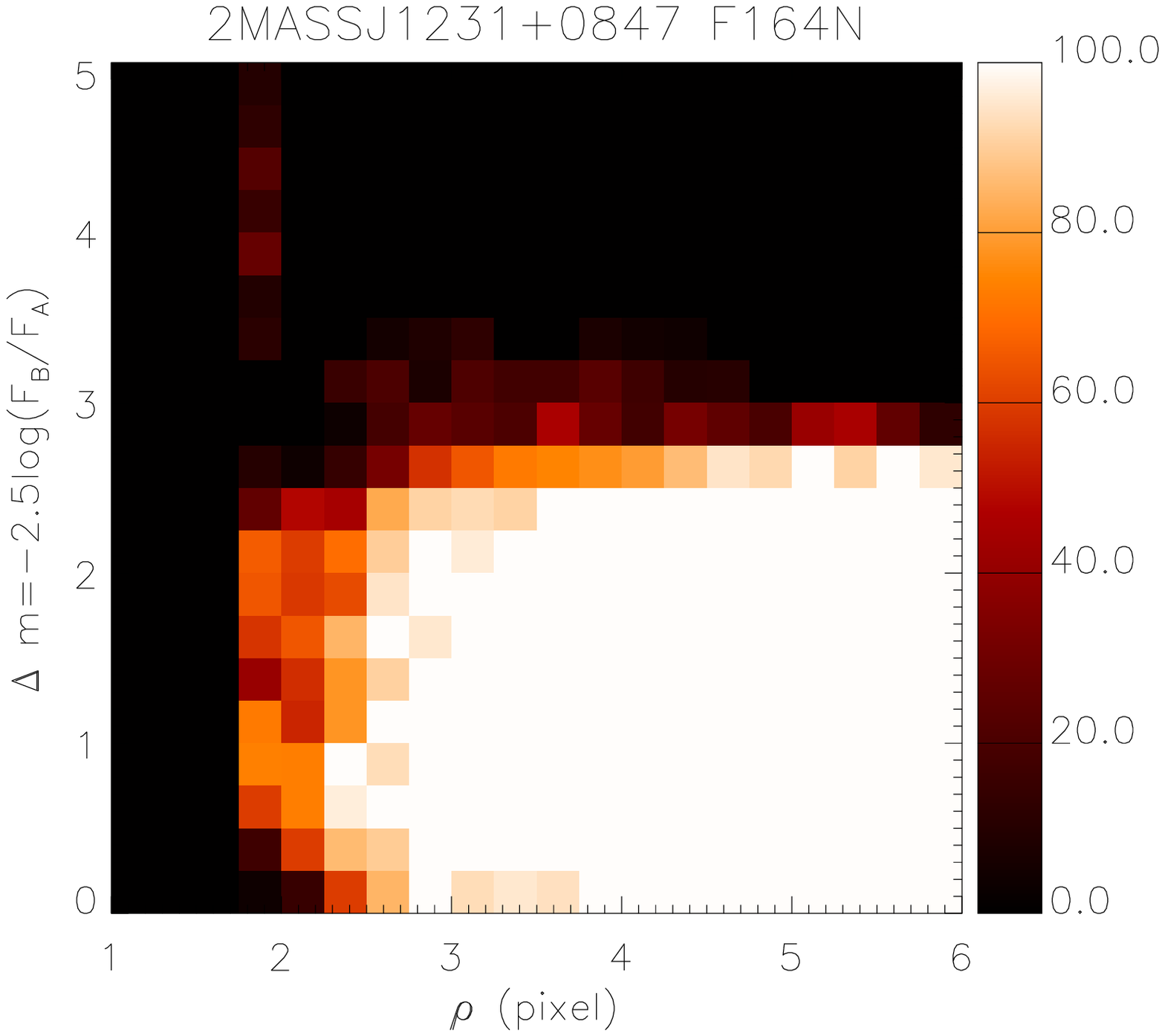}\\
\caption{Samples of selection probabilities based on Monte Carlo simulations described in the text, showing 2MASSJ1754+1649 and SDSS J1628+2308 in F110W filter,  2MASS J0729$-$3954 and HD3651B in F127M filter, and finally  2MASS J0050$-$3322 and 2MASS J1231+0847 in F164N filter. Detection probabilities are indicated  by color scale.}
\label{simulation}
\end{figure*}

These procedures were done for each source in our sample;  an example is shown in Figure \ref{simulation}. Sensitivity maps in $\Delta$m and separation were determined based on the fraction of implants recovered, and nulling regions with high false positive rates. We find WFC3  is able to detect companions at separations greater than 0.325$\arcsec$ and with $\Delta m_{F110W}$$\leq$2.75 mag,   $\Delta m_{F127M}$$\leq$3.0, $\Delta m_{F139M}$$\leq$2.25 and $\Delta m_{F164N}$$\leq$2.5. Thus,  F127M is the most sensitive filter to detect faint companions both due to better image quality (sharper PSF), and since cool (T, Y) companions tend to have a flux distribution that peaks in F127M.

\section{Analysis}
\label{analysis}

\subsection{Comparison with known mid and late-T dwarfs binary systems}
\label{knownmidandlateT}

\begin{table*}
\scriptsize
\caption{Summary of known mid, late-T dwarf binary systems closer than 20 pc.}     
\label{surveys}          
\begin{tabular}{lcccccccc}  
\tableline\tableline                 
Object                              & Instrument     & $\Delta$$m_{F127M}$   & $\rho$                     &  $\rho$         &     Distance                & Spt. A & Spt. B          &Binary        \\
               			    &                           & (mag)                           &   (AU)                       & (mas)               &       (pc)                       &             &                      &Reference         \\
(1)           			    & (2)                     & (3)                                   & (4)                            & (5)                 & (6)                                  & (7)       & (8)      & (9)    \\
\tableline                                                
2MASS J1534$-$2952$^{a}$  &  {\em HST}  WFPC2        & 0.16$\pm$0.28$^{b}$               & 2.3$\pm$0.5         &140.3$\pm$0.57$^{b}$ &   13.6$\pm$0.2  &   T5     &  T5          &\citet{2003ApJ...586..512B}\\
2MASS J1225$-$2739  	    &  {\em HST}  WFPC2        &1.227$\pm$0.05       & 3.8$\pm$0.1        &282$\pm$5     & 13.4$\pm$0.04  &   T6     &  T8           &\citet{2003ApJ...586..512B}\\
2MASS J1553$-$1532              &  {\em HST} NICMOS       &0.052$\pm$0.02                & 4.2$\pm$0.7        & 349$\pm$5   & 12$\pm$2.0       &   T6.5  &  T7           &\citet{2006ApJS..166..585B}\\
WISE J0458+6634                     &   Keck   NIRC2                   &0.944$\pm$0.09          & 5$\pm$0.4         &510$\pm$20  &  10.5$\pm$1.4    &  T8.5    &  T9           &\citet{2011AJ....142...57G} \\
CFBDSIR J1458+1013             &   Keck  NIRC2                    & 1.721$\pm$0.07       &  2.6$\pm$0.3      & 110$\pm$5   & 23.1$\pm$2.4    &T9.5     & $>$T10   &  \citet{2011ApJ...740..108L}\\
WISE J1217+1626 		   & Keck  NIRC2                        & 2.021$\pm$0.03 &  8.0$\pm$1.3 & 759.2$\pm$3.3 	 &10.5$\pm$1.7 & T9   &Y0      & \citet{2012ApJ...758...57L} \\
 WISE J1711+3500		   &   Keck  NIRC2                      & 2.722$\pm$0.03   & 15.0$\pm$2.0	&780.0$\pm$2.0 &  19.0$\pm$3.0 &T8	& T9.5    &\citet{2012ApJ...758...57L}\\
 \tableline
\end{tabular}
{$^{a}$}{This source was not resolved with WFPC2.}\\
{$^{b}$}{$\Delta$$m$ and $\rho$ measured by Keck LGS AO observations on $K_{s}$ filter  (\citealt{2008ApJ...689..436L}).}
\end{table*}

\begin{figure}
\centering
\includegraphics[width=0.5\textwidth]{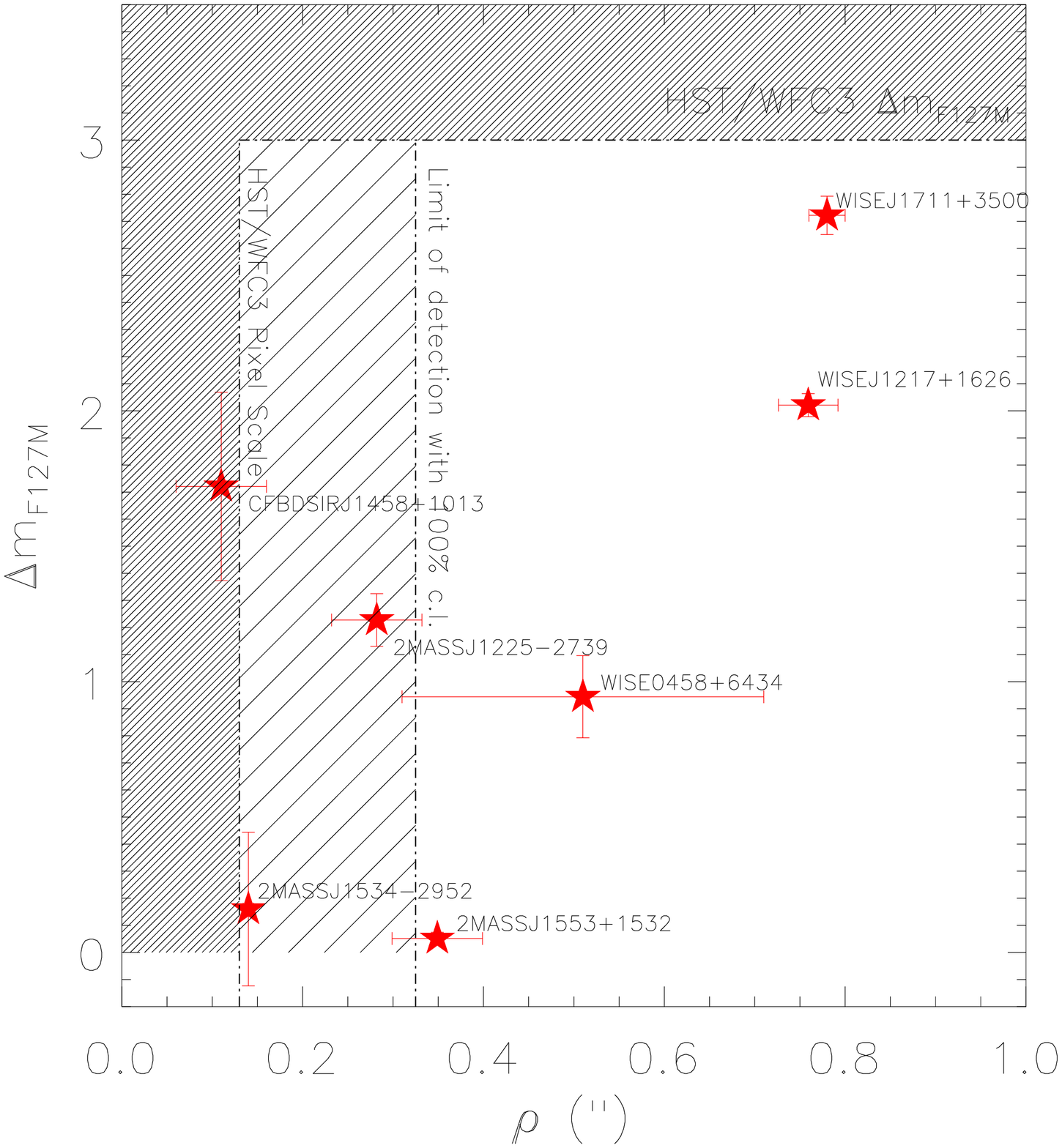}\\
\caption{Mid, late-T dwarf binary systems discovered with Keck II LGS-AO and {\em HST}/NICMOS-WFPC2. Sources are plotted in $\Delta m_{F127M}$ vs. separation and detection limits for the WFC3 program are overlaid.}
\label{deltam_vs_sepF127M}
\end{figure}

High resolution searches with {\em HST}/NICMOS-WFPC2 and Keck/NIRC2 instruments have resulted in the detection of seven mid to late T dwarf binary systems at distances closer than 20 pc (see Table \ref{surveys}). With a total of 90 such sources within that distance limit, the corresponding visual BF is 7.8$^{+7.5}_{-3.9}$$\%$. However, a proper comparison requires a quantification of selection effects. \\

We first calculated the probability of resolving the known mid, late T dwarf binaries in our sample with WFC3. We transformed resolved $J$ magnitudes to F127M using synthetic colors computed from low- resolution near-infrared spectra of L0-T9 dwarfs from the SpeX Prism Spectral Libraries\footnote{We generated a 6$^{th}$ order polynomial to fit synthetic colors for 543 L0-T9 dwarfs, F127M-$J$ = $\sum_{i=0}^{6} a_i$SpT$^i$, where SpT(L0) = 20, SpT(T0) = 30, etc. The fit coefficients $\vec{a}$ = [-9.94976e1, 2.11571e1, -1.84264e0, 8.40307e-2, -2.11748e-3, 2.79814e-5, -1.51800e-7] have a standard deviation of 0.016~mag.}. Figure \ref{deltam_vs_sepF127M} compares separations and relative magnitudes for these systems to our WFC3 sensitivity limits.  Only 2MASS J1553+1532, WISE J0458+6434, WISE J1217+1626 and WISE J1711+3500 are within  the WFC3 limits. 
Multiplying the visual BF with the probabilities $P$($\rho$$\geq$0.325$\arcsec$) and $P$($\Delta$m$_{F127M}$$\le$3 mag), the probability of finding T5+$\ge$T5 dwarf binaries in our sample is 4.4$\%$, which is in agreement with the null binary detection in the studied sample.

\subsection{Inferring the Binary Fraction of Brown Dwarfs for T5+ primary companions.} 
Given the WFC3 pixel scale, the absence of any new discoveries in this sample is 
not wholly unexpected.  Nevertheless, our sample, is the largest containing T5+ sources, so it allows us to more tightly constrain the underlying binary fraction.  To do this, we applied the detection and false positive rate maps computed for each source in the F127M filter to another Monte Carlo simulation 
that determines the probability that each source, if it were a binary, would have been uncovered.  

We generated a large sample (5$\times$10$^6$) of binaries 
by first drawing primary masses from a power-law mass distribution quantified as d$N$/dM $\propto$ M$^{-0.5}$ \citep{2004ApJS..155..191B,2013MNRAS.433..457B}.  Secondary masses were then computed assuming either a flat mass ratio distribution ($P(q \equiv$ M$_2$/M$_1$) $\propto$ constant) or a power-law distribution ($P(q) \propto q^{1.8}$;  \citealt{2007ApJ...668..492A}), imposing a minimum mass of 0.005~M$_{\odot}$.  Adopting a uniform age distribution between 0.1~Gyr and 10~Gyr for the simulated systems, the component masses were converted to bolometric luminosities using the evolutionary models of \citet{2001RvMP...73..719B}, and these transformed into spectral types and absolute $J$ magnitudes using the relations given in \citet{2012ApJS..201...19D}. $J$ magnitudes were then transformed to F127M using synthetic colors computed as above. We then computed relative F127M magnitudes for each of the simulated binaries.  For the orbits, we assigned semimajor axes assuming either a flat ($P(a) \propto$ constant) or lognormal distribution:
\begin{equation}
P(\log{a}) \propto e^{-\left(\frac{\log{a}-0.86}{0.28}\right)^2}
\end{equation}
\citep{2007ApJ...668..492A} where $a$ is in AU and constrained to be $<$25~AU, a limit which encompasses known T dwarf field binaries.
We assumed uniform distributions of mean anomaly, longitude of ascending node, and argument of periapsis, a sin i distribution for inclination, and a uniform distribution of eccentricities over 0 $< e <$ 0.6 based on the analysis of \citet{2011ApJ...733..122D}.
These orbital elements were projected onto the sky and transformed into angular separations at the distance of each system.  We selected only those systems whose primary spectral type was within 1 subtype of the target, and determined the fraction of these that could have been resolved with WFC3 based on our detection and false positive rate maps. We further assumed that companions wider than 0$\farcs$6 could be detected to the sensitivity limit of each image (Table \ref{log}).  We computed fractions for the four possible combinations of mass ratio and separation distributions as described above (see Table \ref{tab:binsim}).  

\begin{figure*}[h!]
\centering
\includegraphics[width=0.7\textwidth]{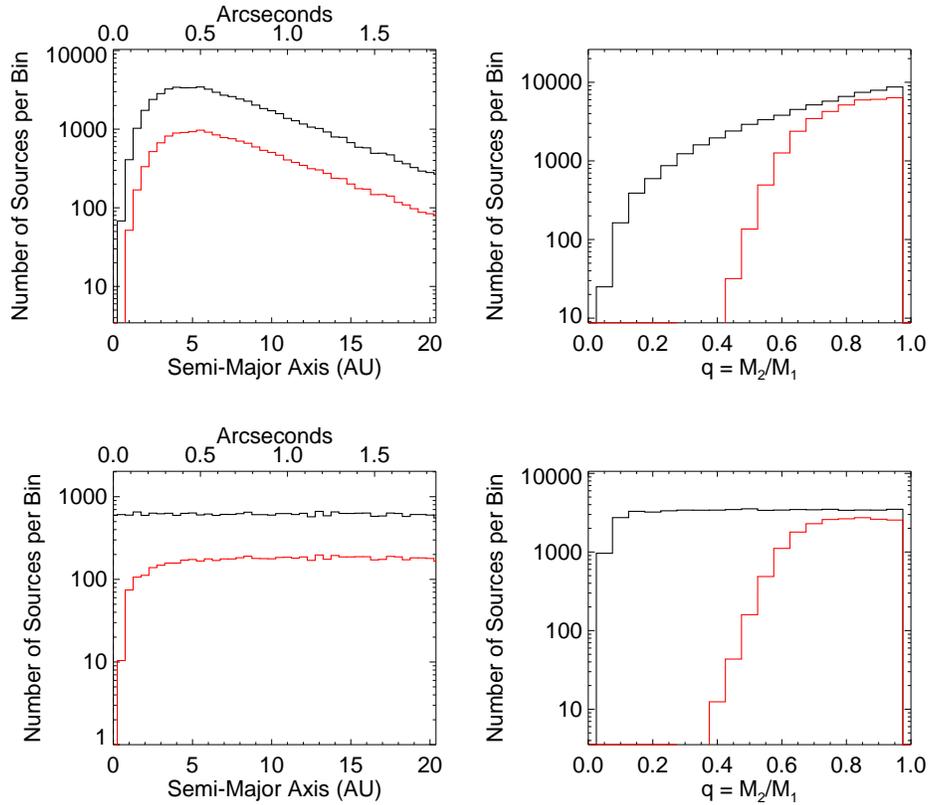}
\caption{Binary detection probability distributions for the T5.5 2MASS~J1828$-$4849 as a function of semi-major axis (left) and mass ratio (right) based on the simulations described in the text. Each panel displays the distributions of input (black lines) and recovered (red lines) systems based on WFC3 selection function for this source.  The top and bottom left panels compare  lognormal and constant input distributions for semi-major axis; the top and bottom right panels compare power-law and constant input distributions in mass ratio, respectively. The resulting total binary recovery rate for this and other sources in our sample are given Table~\ref{tab:binsim}.}
\label{fig:simbf1}
\end{figure*}

Table~\ref{tab:binsim} lists the resulting probabilities of detection, while Figure~\ref{fig:simbf1} illustrates how WFC3 selection effects impact observed distributions of binary separation and mass ratio in the case of 2MASS~J1828$-$4849.  Not surprisingly, both the closest systems ($<$1~AU) and lowest-$q$ systems ($q < 0.6$) are preferentially lost, the latter having the more significant impact on overall recovery rate. 
The most distant targets in our sample have the lowest  detection probabilities and the largest differences in detectability based on the assumed separation distribution, a consequence of the peak of the lognormal distribution falling below angular resolution limits.  Detection probabilities are consistently lower for flat versus power-law mass ratio distributions. 

By adding up the individual source probabilities, we find that if all our targets had companions we should have detected between 13 and 21 binaries, depending on the assumed underlying distribution.  The lack of detections implies a binary fraction upper limit of $<$16 -- $<$25\% assuming a binomial distribution with 95$\%$ of confidence level\footnote{To calculate the confidence limits we use the Clopper-Pearson exact method based on the beta distribution (Brown et al. 2001).}.

\begin{figure*}[h!]
\centering
\includegraphics[width=0.5\textwidth]{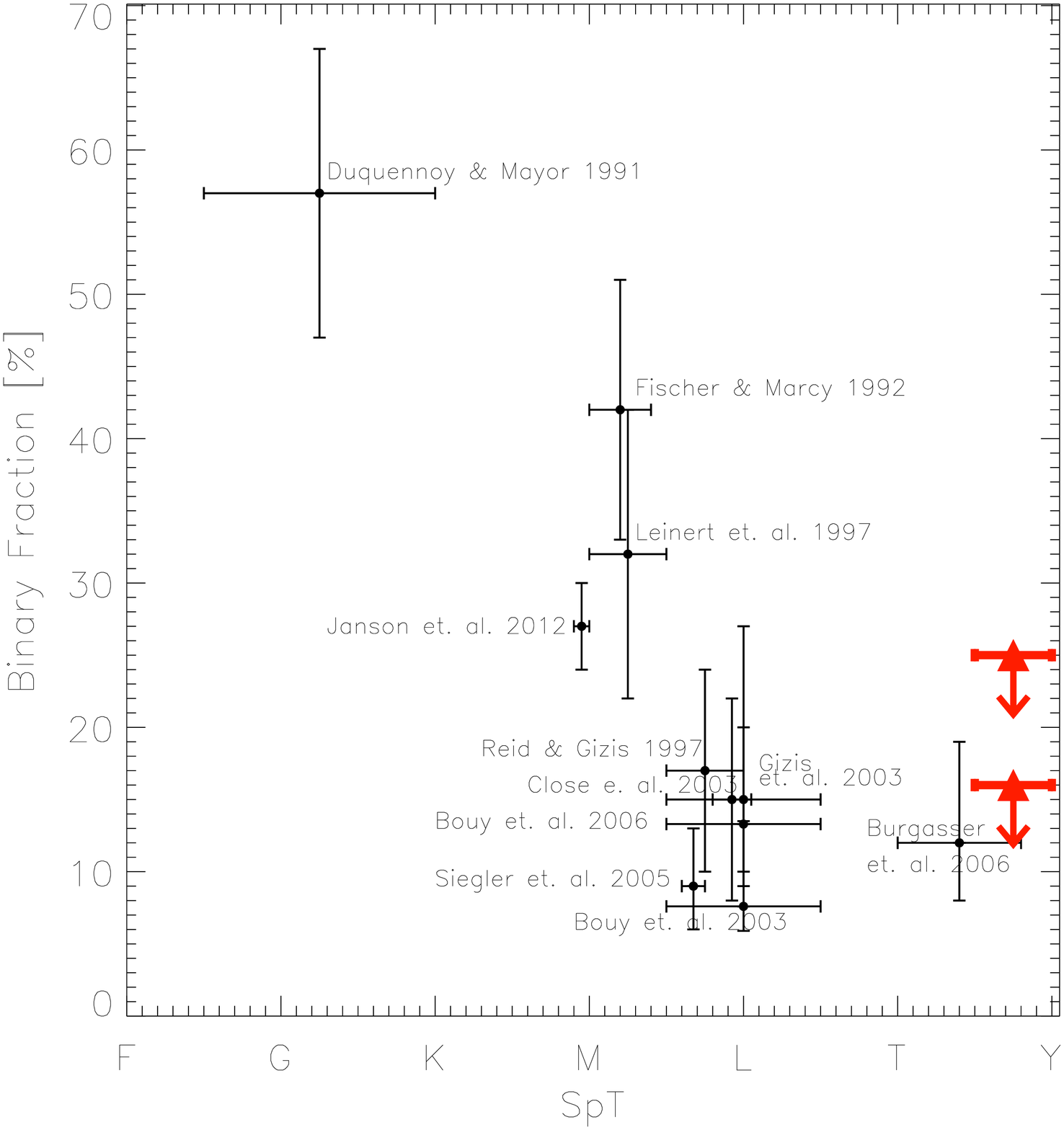}
\caption{Binary frequency as a function of the spectral type in the field and in clusters. The upper limits determined in this work are shown with red triangles.}
\label{bf_vs_spt}
\end{figure*}

 These values are consistent with previous {\em bias-corrected} estimates of the field brown dwarf binary fraction \citep{2003ApJ...586..512B,2006ApJS..166..585B,2007ApJ...668..492A} and supports the hypothesis that multiplicity rates decline with decreasing primary mass into the substellar regime (see Figure \ref{bf_vs_spt}; \citealt{1992ApJ...396..178F}; \citealt{1997AJ....113.2246R}; \citealt{2003AJ....126.1526B}; \citealt{2003ApJ...587..407C}; \citealt{2012ApJ...757..141K}; \citealt{2012MNRAS.419.3115B}). However, our estimates are subject to the same limitations on probing the closely separated ($<$1 AU) binary population as prior imaging programs. The resolving limit of HST and AO imaging coincides with the peak of the brown dwarf binary separation distribution, suggesting that tighter binaries may be plentiful (\citealt{2006ApJS..166..585B}). Alternate detection methods such as RV monitoring (e.g., \citealt{1999AJ....118.2460B}; \citealt{2010ApJ...723..684B}) or spectral blend detection (e.g., \citealt{2010ApJ...710.1142B}) are still needed to determine if the brown dwarf binary 
fraction may in fact be much higher than imaging studies indicate.

\begin{table*}[h!]
\scriptsize
\centering
\caption{Companion Detectability with WFC3.}
\label{tab:binsim}
\begin{tabular}{lcccccc}     
\tableline\tableline  
Name & SpT & Distance & Power-Law q & Power-law q & Flat  q & Flat q  \\
            &         &  (pc)         &  Lognormal a  & Flat a              & Lognormal a  &Flat a  \\
  (1)     &  (2)   &  (3)          &  (4)                    &  (5)                  & (6)                     & (7)   \\
\tableline\tableline  
ULAS J0034$-$0052 & T8.5 & 12.6$\pm$0.6 & 82\% & 84\% & 60\% & 61\% \\
HD 3651B & T7.5 & 11.0$\pm$0.1 & 74\% & 74\% & 45\% & 45\%  \\
2MASS J0050$-$3322 & T7.0 & 8.00$\pm$1.0 & 80\% & 81\% & 51\% & 52\%  \\
SDSS J0325+0425 & T5.5 & 19.0$\pm$2.0 & 69\% & 75\% & 44\% & 47\%  \\
2MASS J0407+1514 & T5.0 & 17.0$\pm$2.0 & 83\% & 86\% & 60\% & 63\%   \\
2MASS J0510$-$4208 & T5.0 & 18.0$\pm$2.0 & 72\% & 79\% & 46\% & 51\%  \\
2MASSI J0727+1710 & T7.0 & 9.10$\pm$0.2 & 88\% & 88\% & 62\% & 62\%  \\
2MASS J0729$-$3954 & T8.0 & 6.00$\pm$1.0 & 94\% & 93\% & 76\% & 76\% \\
2MASS J0741+2351 & T5.0 & 18.0$\pm$2.0 & 75\% & 80\% & 50\% & 52\%  \\
2MASS J0939$-$2448 & T8.0 & 10.0$\pm$2.0 & 82\% & 82\% & 56\% & 56\%  \\
2MASS J1007$-$4555 & T5.0 & 15.0$\pm$2.0 & 77\% & 80\% & 51\% & 53\%  \\
2MASS J1114$-$2618 & T7.5 & 10.0$\pm$2.0 & 83\% & 83\% & 57\% & 56\% \\
2MASS J1231+0847 & T5.5 & 12.0$\pm$1.0 & 76\% & 78\% & 48\% & 48\% \\
ULAS J1238+0953 & T8.5 & 18.5$\pm$4.3 & 70\% & 77\% & 47\% & 53\%  \\
SDSSP J1346$-$0031 & T6.5 & 14.5$\pm$0.5 & 82\% & 87\% & 59\% & 63\% \\
SDSS J1504+1027 & T7.0 & 15.9$\pm$2.5 & 69\% & 75\% & 42\% & 46\% \\
SDSS J1628+2308 & T7.0 & 14.0$\pm$4.0 & 78\% & 82\% & 52\% & 56\%  \\
2MASS J1754+1649 & T5.0 & 14.3$\pm$2.4 & 74\% & 77\% & 46\% & 48\%  \\
SDSS J1758+4633 & T6.5 & 12.0$\pm$2.0 & 76\% & 77\% & 48\% & 48\%  \\
2MASS J1828$-$4849 & T5.5 & 11.0$\pm$1.0 & 69\% & 73\% & 40\% & 43\%  \\
2MASS J1901+4718 & T5.0 & 15.0$\pm$2.0 & 75\% & 80\% & 48\% & 52\%  \\
SDSS J2124+0100 & T5.0 & 18.0$\pm$2.0 & 74\% & 79\% & 50\% & 52\%  \\
2MASS J2154+5942 & T5.0 & 10.0$\pm$1.0 & 86\% & 86\% & 60\% & 61\%  \\
2MASS J2237+7228 & T6.0 & 13.0$\pm$2.0 & 79\% & 82\% & 51\% & 55\% \\
2MASS J2331$-$4718 & T5.0 & 13.0$\pm$2.0 & 78\% & 81\% & 50\% & 52\%   \\
2MASS J2359$-$7335 & T6.5 & 12.3$\pm$1.9 & 79\% & 80\% & 51\% & 52\% \\
\cline{1-7}
Total Expected & &  & 20.2  & 21.0  & 13.5 & 14.0  \\
T5+ Dwarf Binary fraction$^{a}$ & &  & $<$17$\%$ & $<$16$\%$ & $<$25$\%$ & $<$23$\%$ \\
\tableline               
\end{tabular}
\begin{minipage}{13cm}
{$^{a}$}{with 95$\%$ of confidence level.}\\
\end{minipage}
\end{table*}

\section{Conclusions}
\label{conclusions}
 We have analysed data obtained in two imaging survey of 34 BDs with HST/WFC3. The sample comprises 8 L and T dwarfs that we have used to study color-color and color-SpT relations, and 26 mid- to late T dwarfs employed in a search for $\ge$T5 dwarfs companions. Only one previously identified widely-separated system was recovered: 2MASS J1520-4422AB. PSF-fitting uncovered no new close companions to mid-late T sources in our sample.\\

Based on  Monte Carlo simulation we should have been able to detect faint objects at separations $\ge$0.325$\arcsec$ and with $\Delta$$m_{F127M}$ $\leq$ 3.0. Our failure to detect such companions implies a low binary fraction or a significant population of tight binaries. We determined the fraction of binaries that would have been detected around each source based on assumed separation and mass ratio distributions and all possible orientations of these systems. Due to the WFC3 separation limit that makes the null detection of these sources, we infer an upper limit for the binary fraction of  $<$16 -- $<$25\,$\%$, depending of the underlying mass ratio distribution. Comparing with previous BD binary surveys made with HST, we can conclude that WFC3 is more sensitive to cool companions than NICMOS and WFPC2 but its lower resolution makes it poorly suited for typically tight brown dwarf binary systems.

\acknowledgments
Based on observations made with the NASA/ESA Hubble Space Telescope, obtained from the Data Archive at the Space Telescope Science Institute, which is operated by the Association of Universities for Research in Astronomy, Inc., under NASA contract NAS 5-26555. These observations are associated with programs GO-11631 and GO-11666. Support for these programs were provided by NASA through a grant from the Space Telescope Science Institute, which is operated by the Association of Universities for Research in Astronomy, Inc., under NASA contract NAS 5-26555. 
We acknowledge that this project has benefited from contributions by T. Dupuy, J. Faherty, M. Ireland and M. Liu who were co-investigators and helped develop the original HST proposals, GO 11631 and GO 11666. They provided proprietary information on target selection based on ongoing surveys with ground-based laser guide-star adaptive optics (Liu, Dupuy, and Ireland) and astrometry (J. Faherty).
This research has made possible thanks to an international grant from the Spanish Industry Ministry. This research has been supported by  the Spanish Virtual Observatory (http://svo.cab.inta-csic.es), project  funded by MICINN / MINECO through grants AyA2008-02156, AyA2011-24052. This research has benefitted from the M, L, T, and Y dwarf compendium housed at DwarfArchives.org and has benefitted from the SpeX Prism Spectral Libraries, maintained by Adam Burgasser at http://pono.ucsd.edu/~adam/browndwarfs/spexprism. Special thanks to Daniella Bardalez Gagliuffi, Juan Carlos Mu\~{n}oz, Julia Alfonso Garz\'on and Benjamin Montesinos. 
\clearpage
\bibliography{biblibrary}
\clearpage

\appendix
\section{Appendix A}
\label{2MASSJ1520$-$4422AB}
\subsection{2MASSJ1520$-$4422AB}
The only well-resolved target in our sample is the previously identified L dwarf binary 2MASS~J1520$-$4422AB, originally reported by \citet{2007MNRAS.374..445K} and \citet{2007ApJ...658..557B} and found to have an angular separation of 1174$\pm$16~mas at position angle 27$\fdg$1$\pm$0$\fdg$7 (east of north; epoch 2006 April 8 UT)\footnote{In \citet{2007ApJ...658..557B}, the position angle of this binary is reported as 152.9$\fdg$1$\pm$0$\fdg$7, pointing from primary to secondary. However, the authors failed to take into account an image flip in the data, so the actual position angle of the source should have been reported as {\bf 27$\fdg$1.  Our measurement have been also verified in our LDSS3 acquisition images.}} and an estimated projected separation of 22$\pm$2 AU.  Our WFC3 observations yield a separation of 1.20$\pm$0.01$\arcsec$ and PA=29.65$\fdg$$\pm$0.70$\fdg$.

The components of this system are classified L1.5 and L4.5 based on NIR spectroscopy, and to date only a combined-light optical spectrum has been reported \citep{2008MNRAS.383..831P}.  Because the optical spectra of L dwarfs contain a number of diagnostics of age and mass, including H$\alpha$ emission at 6563~{\AA} and Li~I absorption at 6708~{\AA}, we obtained resolved optical spectroscopy of the system using the Low Dispersion Survey Spectrograph (LDSS-3; \citealt{1994PASP..106..983A}) mounted on the Magellan 6.5m Clay Telescope.
Observations were obtained on 2006 May 8 (UT) in clear conditions 
with moderate seeing (0$\farcs$7 at $R$-band). Data acquisition and reduction procedures are identical to those described in \cite{2009AJ....138.1563B}. \\

\begin{figure*}[h!]
\centering
\includegraphics[width=0.65\textwidth]{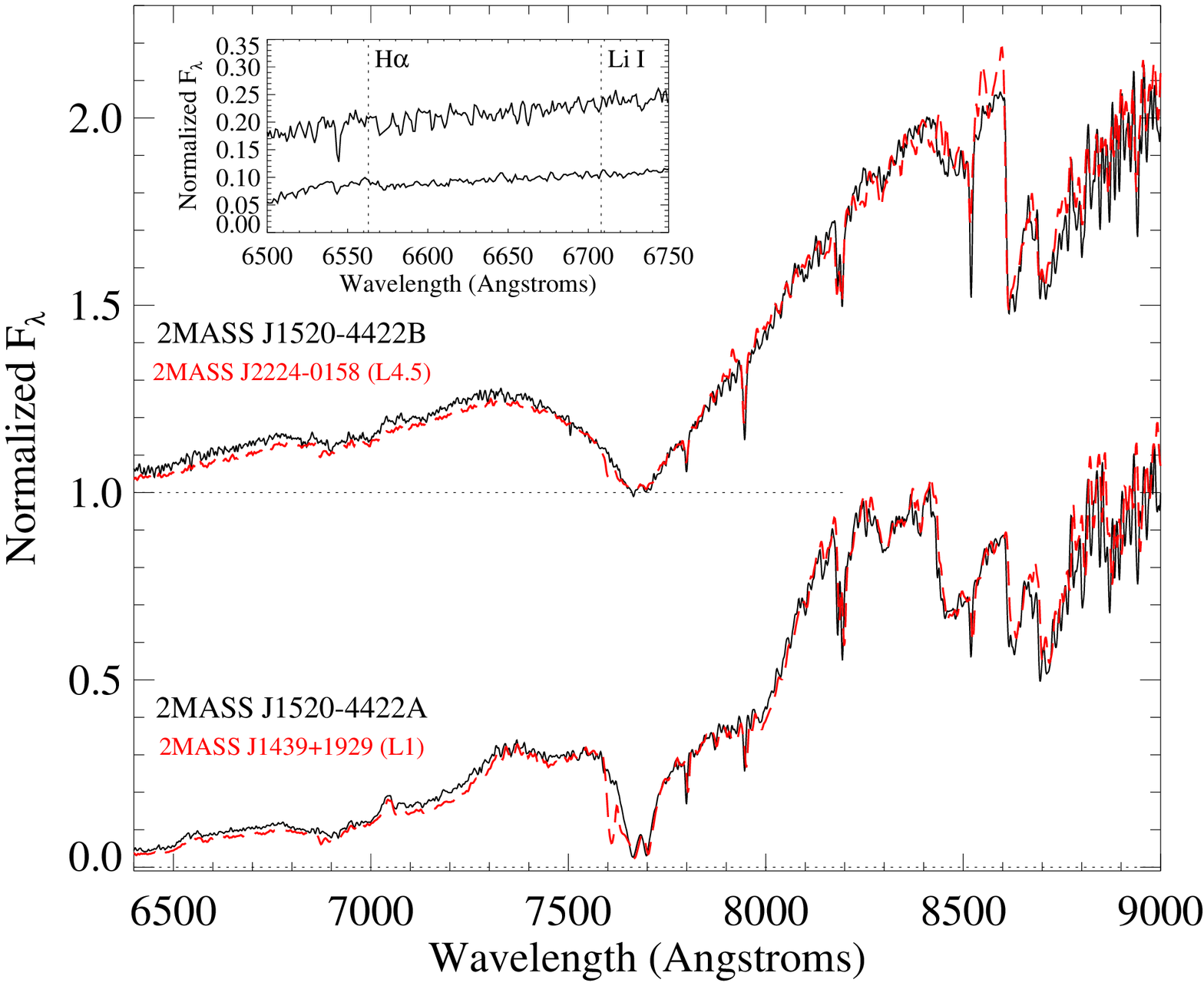}
\caption{Optical spectra of 2MASS~J1520$-$4422AB (solid black lines)
compared to the L1 standard 2MASS J14392836+1929149 \citep{1999ApJ...519..802K} and the
L4.5 2MASS J22244381$-$0158521 (\citealt{2000AJ....120..447K}; red dashed lines). All spectra are gaussian-smoothed to a common resolution of $\lambda/\Delta\lambda$ = 1500 and normalized at 8300~{\AA}, with the L4.5 dwarfs offset by a constant (dotted line).  Note that the comparison spectra have not been corrected for telluric absorption (7150--7300~{\AA}; 7600-7650~{\AA}).  The inset box highlights the 6500--6750~{\AA}, region revealing no evidence of H$\alpha$ emission or Li~I absorption.}
\label{fig:opt1520}
\end{figure*}

Figure~\ref{fig:opt1520} displays the reduced red optical spectra of both components, compared to equivalent data for the 
L1 standard 2MASS J14392836+1929149 \citep{1999ApJ...519..802K} and the
L4.5 2MASS J22244381$-$0158521 \citep{2000AJ....120..447K}.
The overall spectral morphologies between the  2MASS~J1520$-$4422AB components and  templates are in good agreement, confirming the NIR classifications. Note that the 8521~{\AA} Cs~I line in 2MASS~J1520-4422B is considerably stronger than the template, which may reflect slight differences in temperature, metallicity or surface gravity. 

\begin{figure*}[h!]
\centering
\includegraphics[width=0.65\textwidth]{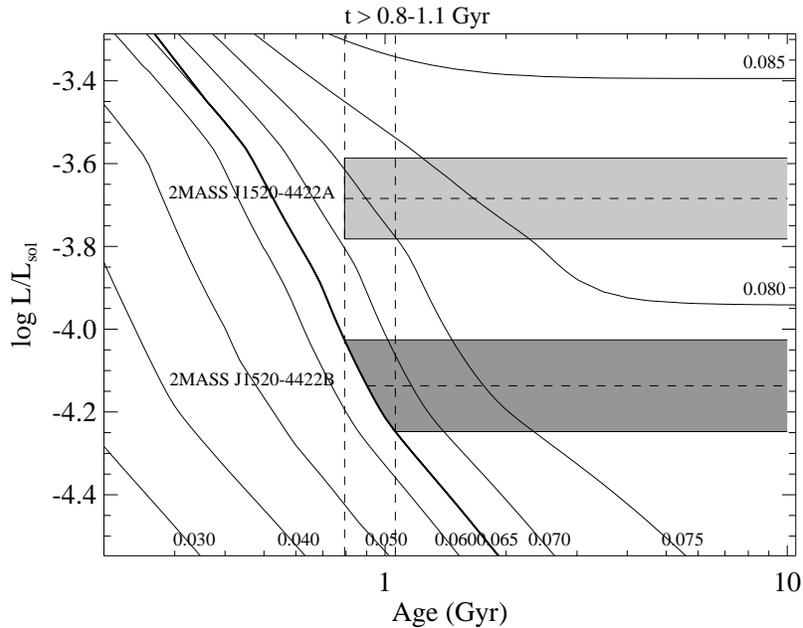}
\caption{Model-dependent age constraints for the 2MASS~J1520$-$4422AB system, based on the absence of Li~I absorption in the component spectra.  Bolometric luminosities as a function of time are shown for various masses (labeled in units of M$_{\odot}$), based on the models of \citet{2001RvMP...73..719B}.
Component luminosities of 2MASS~J1520$-$4422AB (horizontal dashed lines) were estimated from the $M_{bol}$/SpT relation of \citet{2007ApJ...659..655B} and include uncertainties in that relation and component optical classifications (L1$\pm$0.5 and L4.5$\pm$0.5; shaded regions).  Assuming a minimum mass of 0.065~M$_{\odot}$ for both components (thick mass track), we infer a minimum system age of 0.8--1.1~Gyr.}
\label{fig:evol1520}
\end{figure*}

Importantly, neither component shows evidence of H$\alpha$ emission or Li~I absorption.  The latter implies individual masses greater than 0.065~M$_{\odot}$ \citep{1992ApJ...389L..83R,1997ApJ...482..442B} and hence a combined system mass greater than 0.13~M$_{\odot}$.  Transforming the measured spectral types into bolometric luminosities using the relation of \citet{2007ApJ...659..655B} and comparing these to the  evolutionary models of \citet{2001RvMP...73..719B}, we infer a minimum system age of 0.8--1.1~Gyr for 2MASS~J1520$-$4422AB and a minimum primary mass of 0.07~M$_{\odot}$ (Figure~\ref{fig:evol1520}).  This system appears to be a fairly normal, inactive field binary with component masses around the hydrogen burning mass limit.  

\clearpage

\section{Appendix B}
\label{DENISJ1013$-$7842}
\subsection{DENISJ1013$-$7842}
\begin{figure*}[h!]
\centering
\includegraphics[width=0.7\textwidth]{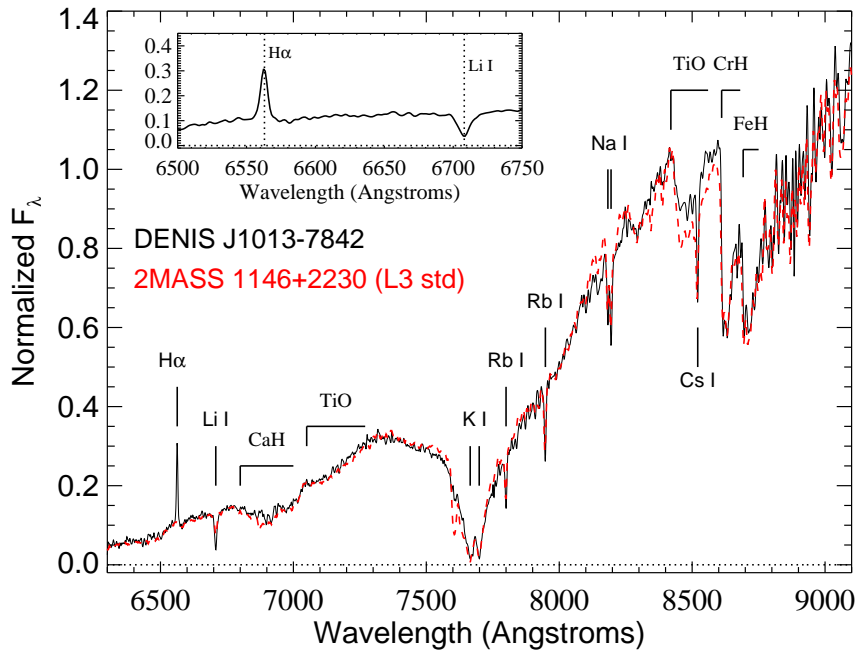}
\caption{Optical spectrum of DENIS~J1013$-$7842 (solid black line) compared to the L3 standard 2MASSW~J1146345+223053 \citep{1999ApJ...519..802K}. Both spectra are gaussian-smoothed to a common resolution of $\lambda/\Delta\lambda$ = 1500 and normalized at 8400~{\AA}.  Note that the comparison spectrum has not been corrected for telluric absorption.  The inset box highlights the 6500--6750~{\AA} region showing strong H$\alpha$ emission and Li~I absorption.}
\label{fig:opt1013}
\end{figure*}

A new source reported here is DENIS~J1013$-$7842, identified in a search for nearby, young, very low-mass objects in the southern sky with DENIS (Looper et al., in prep.).  We obtained an optical spectrum of this source with Magellan/LDSS-3 on 2007 May 8 (UT) in clear conditions with 1$\farcs$3 seeing, using the identical configuration as described above but with the slit aligned with the parallactic angle.  Two exposures of 1500~s were obtained.  The telluric-corrected spectrum is shown in Figure~\ref{fig:opt1013} and compared to that of the L3 optical standard 2MASSW~J1146345+223053 \citep{1999ApJ...519..802K}.  The spectra are nearly identical from 6300--9000~{\AA}, with the exception of DENIS~J1013$-$7842 having pronounced H$\alpha$ emission and somewhat weaker TiO absorption at 8500~{\AA}.  The H$\alpha$ emission is particularly strong, with an equivalent width (EW) = 10.3$\pm$0.2~{\AA}.  Using the $\chi$ formalism of \citet{2004PASP..116.1105W} with a $\chi$ value computed from \citet{2008ApJ...684.1390R} assuming T$_{eff}$ = 1950~K \citep{2004AJ....127.2948V}, we estimate $\log_{10}L_{H\alpha}/L_{bol}$ = $-$5.12$\pm$0.15 for DENIS~J1013$-$7842, consistent with trends among (rare) active early- and mid-type L dwarfs \citep{2007AJ....133.2258S}.  The spectrum of this source also shows strong Li~I absorption (EW = $-$5.8$\pm$0.2), indicating that it is a brown dwarf with M $<$ 0.065~M$_{\odot}$ and age $\lesssim$750~Myr \citep{2001RvMP...73..719B}.  There is no evidence of low surface gravity spectral features in this spectrum, however, so this source is likely to be at least 100-300~Myr old (\citealt{2008ApJ...689.1295K}; \citealt{2009AJ....137.3345C}; \citealt{2010A&A...517A..53M}).

\clearpage

\section{Appendix C}
\label{2MASSJ2237+7228}
\subsection{2MASSJ2237+7228}

One of our HST targets is the previously unreported T dwarf  2MASS J2237+7228.  This source was uncovered by \citet{2007AJ....134.1162L} in a search of the 2MASS survey for mid- and late-type T dwarfs, but at the time of that paper's publication suitable spectral data were unavailable to verify its nature.

\begin{figure*}[h!]
\centering
\includegraphics[width=0.5\textwidth]{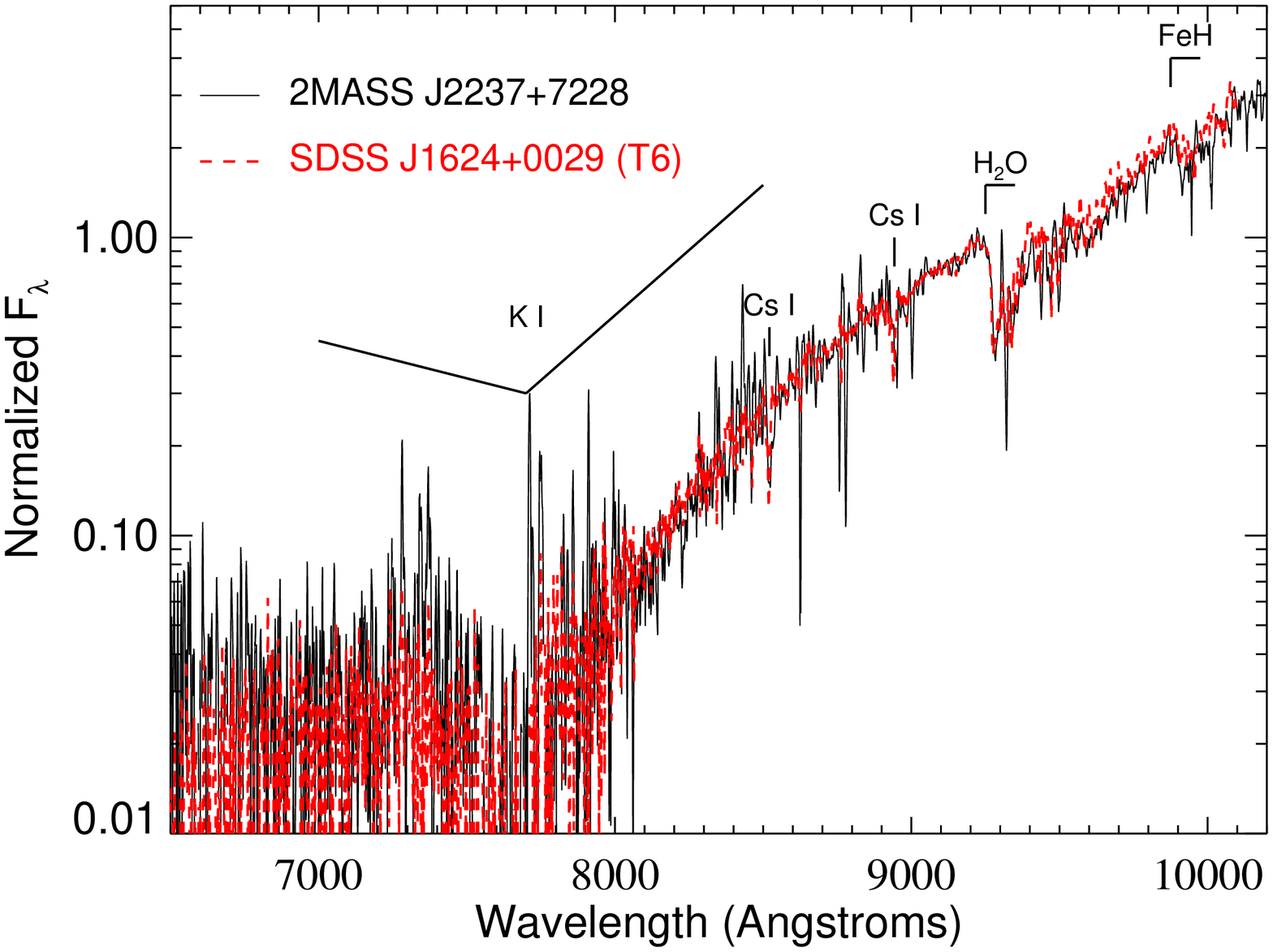}\includegraphics[width=0.5\textwidth]{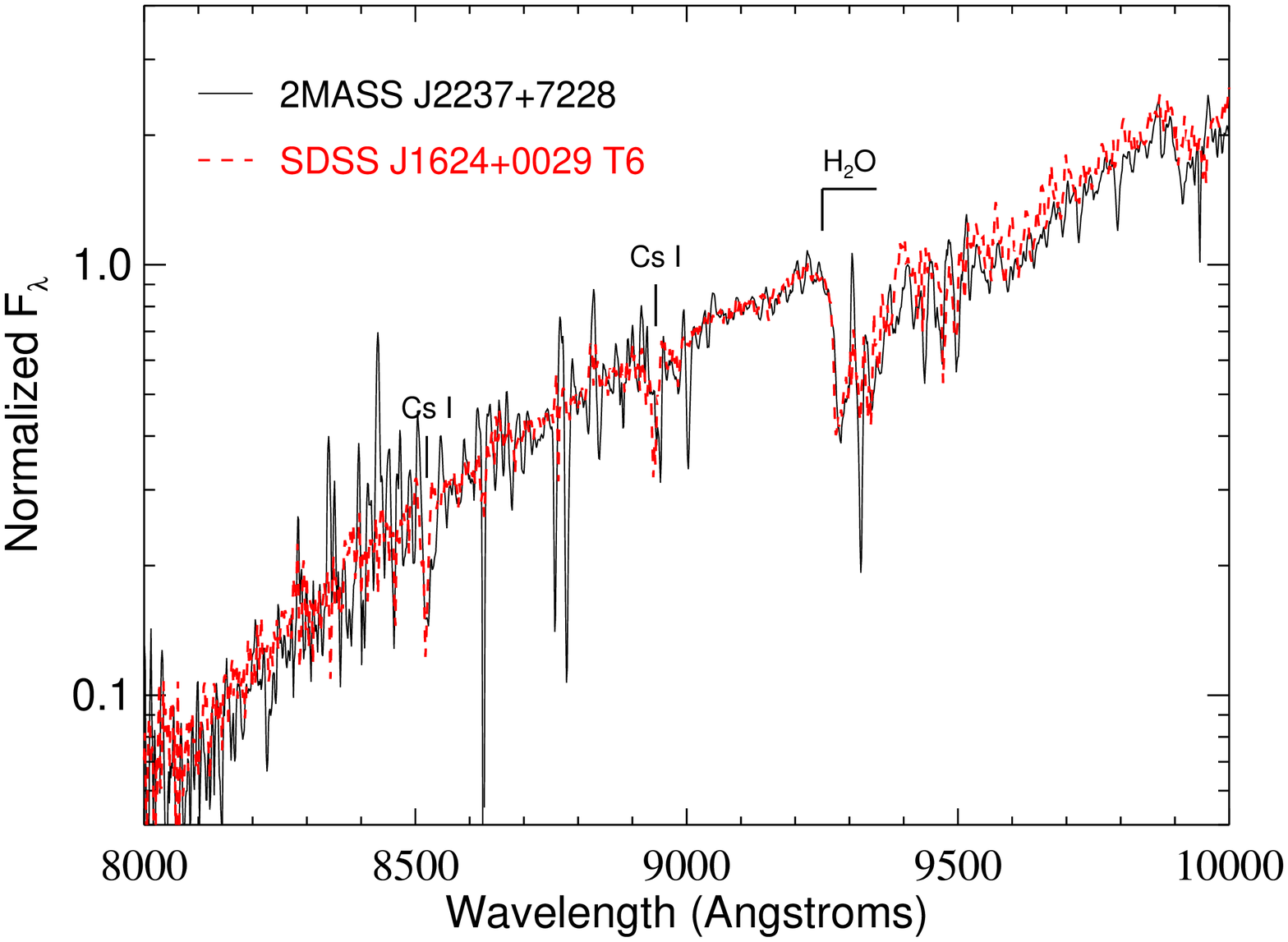}
\caption{Optical spectrum of 2MASS J2237+7228 obtained with Subaru/FOCAS (black line) compared to equivalent data for the T6 spectral standard SDSS J1624+0029 (red line; data from \citealt{2000ApJ...533L.155L}). Both spectra are normalized at 9500~{\AA} and plotted on a logarithmic vertical scale.  Primary absorption features in the red optical spectra of T dwarfs are indicated.}
\label{fig:opt2237}
\end{figure*}

Optical spectral data of 2MASS J2237+7228 were obtained with the Subaru 8m Faint Object Camera and Spectrograph (FOCAS) instrument \citep{2002PASJ...54..819K} on 20 August 2007 (UT) in clear conditions with moderate humidity and light winds. A single 3600~s exposure of the target was obtained with the 0$\farcs$5 longslit, 150~line/mm grating blazed at 6500~{\AA} and SO58 order-blocking filter, providing 5860--10270~{\AA} spectroscopy at a resolution $\lambda/\Delta\lambda$ = 400 and dispersion of 1.3~{\AA}/pixel. A standard flux calibrator from \citet{1994PASP..106..566H} was also observed along with flat field and arc lamps.  Data were reduced using the FOCAS reduction pipeline in IRAF\footnote{Image Reduction and Analysis Facility; \citet{1986SPIE..627..733T}.}; no telluric correction was applied to the data.  Figure~\ref{fig:opt2237} displays the reduced spectrum compared to equivalent data for the T6 dwarf  SDSSp J162414.37+002915.6 \citep{1999ApJ...522L..61S,2000ApJ...533L.155L}, which is an excellent match.  We therefore nominally assign an optical classification of T6 for this source.  

2MASS J2237+7228 is also detected in the WISE survey ($W2$ = 13.62$\pm$0.04, $W1-W2$ = 2.06$\pm$0.07), and comparison of 2MASS and WISE coordinates separated by over a decade indicates a modest proper motion: $\mu_{\alpha}\cos{\delta}$ = $-$73$\pm$2~mas/yr and $\mu_{\delta}$ = $-$116$\pm$2~mas/yr. At the estimated 13$\pm$2~pc distance\footnote{This estimate is based on the 2MASS $J$-band magnitude of the source ($J$ = 15.76$\pm$0.07) and the absolute-magnitude/spectral type relation from \citet{2008ApJ...685.1183L}, assuming an $\pm$0.5 uncertainty on the subtype.} of 2MASSJ2237+7228, this implies a tangential velocity of only 8.3$\pm$1.7~km/s, one of the smallest such motions reported for a T dwarf \citep{2009AJ....137....1F}.

%\bibliography{biblibrary}

%\clearpage

\end{document}